\documentclass[reqno]{amsart}
\usepackage{graphicx}
\usepackage{epstopdf,epsfig}
\usepackage{amsmath} 
\usepackage{amssymb}
\usepackage{mathrsfs}
\usepackage{cleveref}

\usepackage{verbatim}
\usepackage{subfig}
\newcommand{\Rey}{Re}

\numberwithin{equation}{section}

\usepackage{todonotes}

\begin{document}

\title{Reduced-Order Modelling and Closed-Loop Control of the Cylinder Wake}

\author[T. Vojkovi{\'c}]{Tea Vojkovi{\'c}}
\address{Delft Center for Systems and Control \& Laboratory for Aero and Hydrodynamics, Delft University of Technology, The Netherlands}
\email{t.vojkovic@tudelft.nl}

\author[D. Boskos]{Dimitris Boskos*}
\address{Delft Center for Systems and Control, Delft University of Technology, The Netherlands}
\email{a.j.buchner@tudelft.nl}

\author[A.-J. Bucher]{Abel-John Buchner*}
\address{Laboratory for Aero and Hydrodynamics, Delft University of Technology, The Netherlands}
\email{d.boskos@tudelft.nl}

\begin{abstract}
We present a model-based approach for the closed-loop control of vortex shedding in the cylinder wake. The control objective is to suppress the unsteadiness of the flow, which arises at a critical Reynolds number $Re_c$ through a supercritical Hopf bifurcation. In the vicinity of $Re_c$ the flow is well described by a forced Stuart-Landau equation derived via a global weakly nonlinear analysis~\cite{sipp2007global, sipp2012open}. This Stuart-Landau equation governs the evolution of the amplitude $A$ of the global mode on the slow time scale. 
In this paper, we generalize the approach from~\cite{sipp2012open}, which considers a fixed-amplitude harmonic forcing, by allowing the 
forcing amplitude $E'$ to vary on the slow time scale.
This enables the design of closed-loop controllers for multiple surrogate Stuart-Landau models, which we obtain for different classes of forcing frequencies.
When these frequencies are near the global mode oscillation frequency at $\Rey_c$, we can bring both $A$ and $E'$ to zero, which fully suppresses the unsteady part of the flow.
We also show that near this frequency, the optimal forcing structure is in the direction of the adjoint global mode. Assuming partial velocity measurements of the flow,
we design an output-feedback control law that stabilizes the flow. 
The approach hinges on a model predictive controller for the surrogate model, which exploits the full-order model measurements to determine the necessary forcing amplitudes while respecting the modelling constraints.
We achieve suppression of the wake oscillations with spatially dense volume forcing and two-point velocity
measurement at $\Rey=50$.

\end{abstract}


\thanks{A-J. Buchner was supported by the Netherlands Organisation for Scientific Research (NWO), under VENI project number 18176.}
\thanks{*These authors contributed equally to this work.}

\maketitle

\section{Introduction}

Instabilities occurring in flows with oscillator-like behaviour are characterized by self-excited oscillations of distinct frequency~\cite{huerre1998hydrodynamic}. Examples of such instabilities include periodic vortex shedding associated with flow separation, for example in the wake of cylinders or within open cavities~\cite{sipp2007global}. Spatial oscillation of shock waves or shock buffeting~\cite{sartor2015stability} in transonic flows around airfoils is another example. It is of great practical significance to suppress such instabilities, as they cause force and kinematic fluctuations which can, depending on the application context, lead to reduced operational efficiency, induced fatigue and increase in maintenance costs.

The suppression or alleviation of flow instabilities can be achieved via either passive or active flow control techniques. Passive control techniques typically involve modifications of the geometry and, since designed for a specific operational point, can have detrimental effects at off-design conditions. These limitations can be overcome by incorporating active flow control, which can be implemented using either an open- or closed-loop approach. Unlike open-loop control, in which the control input is predefined, closed-loop control uses flow sensor measurements to determine the control action. This makes it capable of adjusting to changing flow conditions. We can distinguish between two types of closed-loop control laws: model-based and model-free control laws. Examples of the latter are proportional-integral-derivative (PID) controllers~\cite{park1994feedback, zhang2005closed, illingworth2014active}, fuzzy logic controllers~\cite{cohen2003fuzzy}, adaptive filters~\cite{kegerise2002adaptive, kestens1998active}, and machine learning-based model-free controllers~\cite{gautier2015closed}. Model-based approaches require flow models which predict the evolution of the flow for a given control input. Such models can be classified  into the following three groups according to \cite{wiener1948cybernetics}: (i) high fidelity models based on first principles such as the discretized Navier–Stokes equations, (ii) low-dimensional models, which approximate the full state dynamics of the underlying physics, and (iii) input-output models which lack the connection to the full-state space (e.g. neural networks). Due to their high dimensionality, high-fidelity models are expensive to simulate. This makes them unsuitable for the design of real-time closed-loop controllers acting on the timescale associated with the instability growth. Therefore, to achieve real-time model-based closed-loop control, low-dimensional models that fall into categories (ii) and (iii) are typically needed. Unlike the input-output models which lack connection with the underlying physics, low-dimensional models which approximate the full state dynamics predict the evolution of the most dominant patterns in the flow (e.g. orthogonal decomposition (POD)–Galerkin models~\cite{noack2003hierarchy}).

In this work, we consider laminar flow around a cylinder exhibiting two-dimensional vortex shedding, which serves as a benchmark for oscillatory flows. Commonly applied actuations for the stabilization of a cylinder wake include suction/blowing on the surface of the cylinder~\cite{park1994feedback, illingworth2014active}, oscillation and/or rotation of the cylinder~\cite{siegel2003feedback, thiria2006wake}, and volume forces~\cite{gerhard2003model, aleksic2014need} such as Lorentz forces in an incompressible electrically conducting viscous flow~\cite{chen2005active}. 

Flow models range from high-dimensional discretized Navier-Stokes and Ginzburg-Landau equation~\cite{roussopoulos1996nonlinear, cohen2005closed} to various low-dimensional models such as linear vortex models~\cite{protas2004linear}, state-space models obtained with Eigensystem Realization Algorithm (ERA)~\cite{illingworth2016model}, and POD-Galerkin models~\cite{gerhard2003model, bergmann2005optimal, aleksic2014need}. 
Various approaches have been proposed to design optimal closed-loop controllers for the linearized discretized Navier-Stokes describing two-dimensional flow past a cylinder. \cite{carini2015feedback} used the minimum control energy (MCE) algorithm proposed by~\cite{bewley2007minimal} while in~\cite{jin2022resolvent} a resolvent-based algorithm in combination with $\mathcal{H}_2$ optimal control tools was proposed. 
On the other hand, the linear low-dimensional models from~\cite{protas2004linear} and~\cite{illingworth2016model} are suitable for classical linear control design including pole placement methods, Linear Quadratic Regulator (LQR) and Linear Quadratic Gaussian (LQG) control, $\mathcal{H}_2$- and $\mathcal{H}_{\infty}$-based methods.
For nonlinear POD-Galerkin models, several nonlinear control approaches have been applied, such as optimal control~\cite{bergmann2005optimal}, energy-based control~\cite{gerhard2003model}, and Model Predictive Control (MPC)~\cite{aleksic2014need}. 

In this work, we construct a nonlinear low-dimensional model of the cylinder wake, which has the form of a forced Stuart-Landau equation and is suitable for closed-loop control design.
For this, we utilize the global weakly nonlinear analysis of the incompressible Navier-Stokes equations introduced in~\cite{sipp2007global} for uncontrolled flow around a cylinder and flow past an open cavity, and later applied to other cases of flows around bluff bodies such as an axisymmetric disc~\cite{meliga2009global} and a spring-mounted cylinder~\cite{meliga2011asymptotic}. In these flows, self-sustained oscillations emerge from stationary conditions if a flow parameter is varied through a supercritical Hopf bifurcation. Therefore, they are characterized by a Jacobian which displays an unstable eigenvalue pair~\cite{sipp2010dynamics}, with the associated structure referred to as the global mode. The global weakly nonlinear analysis relies on the assumption that around the bifurcation point the amplitude $A$ of the global mode evolves on a time scale slower than the flow oscillations. For unforced flows, the evolution of $A$ is governed by the Stuart-Landau equation. This approach was generalized in~\cite{sipp2012open}, which considers a harmonic volume forcing of a cavity flow, characterized by a constant amplitude $E'$, forcing frequency $\omega_f$ and spatial structure $\mathbf{f}_E$. The corresponding forced counterparts of the Stuart-Landau equation were derived based on the scalings used in~\cite{fauve2009}. This model was used to determine actuation suitable for open-loop control. 
It was shown that the flow is at its steady state when both $A$ and $E$ are zero. Thus, the open-loop control strategy, which assumes time-invariant (non-zero) forcing amplitude $E'$, can suppress the unsteadiness of the global mode but insodoing also introduces oscillations in the flow at the frequency of the applied forcing and so does not achieve steady flow
conditions. 

Here, to enable the design of closed-loop controllers, we further generalize the global weakly nonlinear analysis from~\cite{sipp2007global, sipp2012open} by considering a harmonic volume forcing with a slowly-varying amplitude $E'$. As in~\cite{sipp2012open}, for multiple classes of forcing frequencies, we derive corresponding Stuart-Landau equations. These Stuart-Landau equations serve as forced reduced-order models of the flow and are equipped with mappings from their state and input spaces into the corresponding spaces of the full flow. The Stuart-Landau models are parameterized by the Reynolds number $\Rey$ and their coefficients are obtained by just solving a set of steady operator equations, without resorting to computationally intensive unsteady flow simulations. Furthermore, the constructed reduced-order models are suitable for closed-loop controller design due to the slow time variation in the forcing amplitude $E'$. We show that choosing a forcing frequency near the frequency of the global mode at the critical Reynolds number allows us to design a feedback law that brings both $A$ and $E'$ to zero. Thus, we fully suppress the unsteadiness of the flow while minimizing the control effort.
We also introduce an alternative data-driven approach to calculate the coefficients of the Stuart-Landau model. This approach allows us to achieve a higher accuracy in approximating the true flow, but its construction requires unsteady flow simulations at each parameter value of interest. We compare the performance of the models based on the associated control effort.
We furthermore present a methodology for designing an output feedback law from velocity measurements. Using model predictive control, a suitable time-sequence of forcing amplitudes is determined based on the surrogate model. This approach allows us to include soft constraints on the rate of change of the forcing amplitude and thus comply with our modelling requirement of small temporal gradients of $E'$. Although this paper addresses the cylinder wake, the presented methodology can be applied to any flow which can be described by the incompressible Navier-Stokes equation and which loses stability through a supercritical Hopf bifurcation.

We apply our methodology to the reduced-order modelling and control of a cylinder wake at $\Rey=50$. 
For this flow, we present an efficient evaluation of the reduced state from the more realistic case of limited full state measurements, derive and discuss the optimal spatial structure of the volumetric forcing term. We furthermore demonstrate, using both the coefficients derived from the weakly nonlinear analysis and from our data-driven approach, successful suppression of the wake oscillations based on a spatially dense, localized forcing structure and two-point velocity measurement.

The paper is organized as follows. First, we introduce our problem and set the objectives of the paper in Section~\ref{sec:2}. Section~\ref{sec:WNA} addresses our generalization of the weakly nonlinear analysis and derivation of the parametric reduced-order model in the form of a forced Stuart-Landau equation. In Section~\ref{sec:control}, we use this reduced-order model to design a closed-loop controller for the flow. Section~\ref{sec:numericalsetup} describes the numerical setup used for calculating the coefficients of the Stuart-Landau model. Finally, we apply our  
reduced-order modelling and closed-loop control approach to the flow around a cylinder at $Re = 50$ in Section~\ref{sec:results}.

\section{Control of the Flow Over a Cylinder}\label{sec:2}

We consider a two-dimensional flow with freestream velocity $u_{\infty}$ around a cylinder of diameter $D$ forced with a volume force $\mathbf{f}(t)$. This flow is described by the incompressible Navier-Stokes equations, which we consider in the non-dimensionalized form
\begin{subequations} \label{eq:NS}
 \begin{align} 
        \frac{\partial{\mathbf{u}}}{\partial t}&=-\nabla \mathbf{u}\cdot \mathbf{u}-\nabla p+ \frac{1}{Re}\Delta \mathbf{u}+\mathbf{f}(t) \\ 
        \nabla \cdot \mathbf{u}&=0.
    \end{align} 
\end{subequations}
Here $\mathbf{u}=(u, v)$ represents the velocity vector, $p$ is the pressure, and $\Rey$ is the Reynolds number based on the characteristic quantities $u_{\infty}$ and $D$. 

\subsection{Hopf Bifurcation}

In the absence of forcing, i.e., when $\mathbf{f}(t)=\mathbf{0}$, and for $4<\Rey< \Rey_c$, the flow described by~\eqref{eq:NS} is steady, with a symmetric counter-rotating pair of vortices in the near wake. For Reynolds numbers above a critical value $\Rey_c$, the wake periodically oscillates and vortices are shed alternately from each side of 
the cylinder. The transition from stable steady flow (stable equilibrium) to self-sustained oscillatory flow (attracting limit cycle), which occurs at $\Rey_c$, corresponds to a supercritical Hopf bifurcation. This is a bifurcation in which a stable equilibrium of a dynamical system loses stability as a system parameter crosses a threshold value, and an attracting limit cycle arises. Loss of stability is characterized by a pair of complex conjugate eigenvalues of the Jacobian matrix at the equilibrium that cross the imaginary axis of the complex plane while the rest of the eigenvalues remain in the left-half plane. The associated eigenvector, here referred to as the global mode, thus becomes unstable. For Reynolds numbers slightly above the critical value $\Rey_c$, the long-time behavior of the unforced incompressible Navier-Stokes equations~\eqref{eq:NS} can therefore be described by two distinct solutions: The oscillatory flow, which is an attracting limit cycle, and the steady (base) flow $\mathbf{U}_b$, which is an unstable equilibrium. Note that in this work, we are interested in the regime above $\Rey_c$ where the flow can be confined on a two-dimensional manifold that contains both the unstable equilibrium and the limit cycle. Thus we limit our studies to Reynolds numbers $\Rey<230$, after which further bifurcations occur, as verified experimentally by~\cite{williamson1996three} and through DNS by~\cite{jiang2016three}.

\subsection{Problem Formulation}

The aim of this work is to suppress the vortex shedding, i.e. to stabilize the flow to its steady state $\mathbf{U}_b$, by assigning an appropriate feedback law to the forcing $\mathbf{f}(t)$. 
We assume that only partial flow field measurements are available, which we denote by $\mathbf M$.  
This is justified by the fact that flow measurements are typically obtained locally from a finite number of sensors in real systems. 

For the given problem, the governing equations~\eqref{eq:NS} can be written in the compact form
 \begin{subequations}
 \label{eq:NS_matrix}
 \begin{align}
     \mathscr{E} \frac{\partial{\mathbf{U}}}{\partial t}&=\mathscr{N}(\mathbf{U}, \mathbf{f}, \Rey)\label{eq:NS_matrix1} \\
     \mathbf{M}&=\mathscr{H}(\mathbf{U}),\label{eq:NS_matrix2}
 \end{align}       
 \end{subequations}
where
\begin{equation*}
\setlength{\arraycolsep}{3pt}
\renewcommand{\arraystretch}{1.3}
\mathscr{E} = \left(
\begin{array}{cc}
   \mathscr{I}  &   0  \\
  \displaystyle
   0  &  0 \\
\end{array}  \right),\quad \mathbf{U} = \left(
\begin{array}{c}
   \mathbf{u}   \\
  \displaystyle
   p  \\
\end{array}  \right), \quad \mathscr{N}(\mathbf{U}, \mathbf{f}, \Rey) = \left(
\begin{array}{cc}
  -\nabla \mathbf{u}\cdot \mathbf{u}-\nabla p+ \frac{1}{Re}\Delta \mathbf{u}+\mathbf{f}(t)\\
  \displaystyle
    \nabla \cdot \mathbf{u} \\
\end{array}  \right).
\end{equation*}
Here, $\mathscr{N}$ denotes the Navier-Stokes operator, $\mathscr{E}=\mathscr{P}\mathscr{P}^T$ where $\mathscr{P}$ is the prolongation operator that maps the velocity field $\mathbf{u}$ into $(\mathbf{u},0)^T$, $\mathscr{P}^T$ is the restriction operator mapping $(\mathbf{u},p)^T$ to $\mathbf{u}$, and $\mathscr{I}$ denotes the identity operator. Expression~\eqref{eq:NS_matrix1} is a shorthand for the equations of motion~\eqref{eq:NS} while~\eqref{eq:NS_matrix2} models the measurements $\mathbf{M}$ obtained by the sensors as a function $\mathscr{H}$ of the state $\mathbf{U}$.

\begin{figure}
\centering
\includegraphics[width=\textwidth, trim = 0cm 11.5cm 0cm -0.5cm, clip=true]{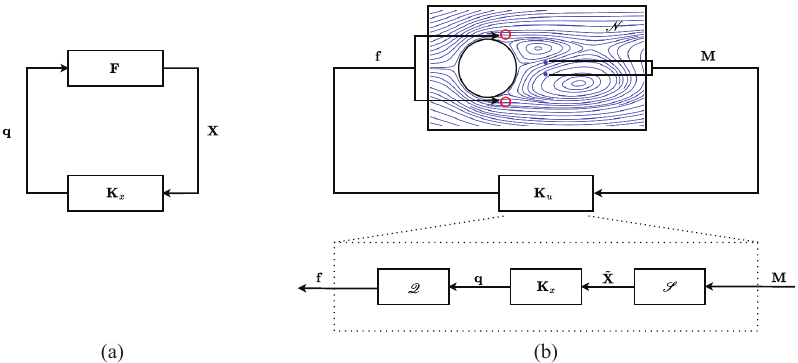}
\caption{Schematic of the closed-loop architecture implemented here to control the cylinder wake. (a) State-feedback loop for the surrogate model. (b) Output-feedback loop for the forced incompressible Navier-Stokes equations (full-order plant).}
\end{figure}

The design of model-based control laws that hinge on real-time predictions requires a fast integration of the system's dynamics.
The numerical approximation of the forced Navier-Stokes equations~\eqref{eq:NS_matrix} is however typically reliant on a fine spatial discretization, which yields a high-dimensional model. Designing controllers for high-dimensional nonlinear systems is particularly challenging, especially through model-based online optimization approaches. Therefore, we seek a parametric surrogate model
 \begin{align}\label{eq:rom}
       \frac{d{\mathbf{X}}}{d t}&=\mathbf{F}(\mathbf{X}, \mathbf{q}, \Rey)
\end{align}
with a low dimensional state $\mathbf{X}$, input $\mathbf{q}$, and mappings $\mathscr{G}$, $\mathscr{Q}$ from the state and input space of \eqref{eq:rom} to the corresponding spaces of the original system~\eqref{eq:NS_matrix1}, so that  $\mathbf{U} \approx \mathscr{G}(\mathbf{X})$, where $\mathbf{U}$ is the solution  of~\eqref{eq:NS_matrix1} with input $\mathbf{f}=\mathscr{Q}(\mathbf{q})$ and $\mathbf{X}$ is the corresponding solution of~\eqref{eq:rom}.
%
%
Our goal is to design an appropriate feedback law
\begin{equation*}
    \mathbf{q}=\mathbf{K}_x(\mathbf{X},\Rey)
\end{equation*}
for~\eqref{eq:rom} and refine it into an output-feedback law $\mathbf{f}\equiv \mathbf{K}_u(\mathbf{M})=\mathscr{Q}(\mathbf{K}_x(\widetilde{\mathbf{X}},\Rey))$ for the Navier-Stokes equations~\eqref{eq:NS_matrix}, which brings the flow to its steady state from a broad set of initial conditions. Here the corresponding
state $\tilde{\mathbf{X}}=\mathscr{S}(\mathbf{M})$ of~\eqref{eq:rom} is estimated from the flow measurements $\mathbf{M}$~\eqref{eq:NS_matrix2} of the full-order system.

\section{Weakly Nonlinear Analysis}\label{sec:WNA}
 
In the following, we generalize the weakly nonlinear analysis of the incompressible Navier-Stokes equations \eqref{eq:NS_matrix1} from \cite{sipp2007global,sipp2012open}. 
The analysis is carried out close to their critical point of the first Hopf bifurcation and enables the derivation of a parametric surrogate model in the form of a forced Stuart-Landau equation.
 Although the paper deals with the cylinder wake, its analysis is suitable for any flow described by~\eqref{eq:NS_matrix1} which losses stability through a supercritical Hopf bifurcation.
%
Assuming the Reynolds number $Re$ to be slightly above the critical value $Re_c$, we define the bifurcation parameter 
\begin{equation}\label{eq:reynolds}
    \epsilon:=Re_c^{-1}-Re^{-1}, \quad 0< \epsilon \ll 1.
\end{equation}

The approach hinges on a multiple timescale analysis, which separates the dynamics on the fast timescale $t$  associated with the oscillations on the limit cycle, from the saturation onto the limit cycle, which evolves on the slow timescale $\tau=\epsilon t$.
The flow and pressure field $\mathbf{U}$ is asymptotically expanded as
\begin{equation}\label{eq:expansion}   \mathbf{U}(t)\approx \mathbf{U}_0+\sqrt{\epsilon}\mathbf{U}_1(t,\tau)+\epsilon \mathbf{U}_2(t, \tau)+\epsilon \sqrt{\epsilon} \mathbf{U}_3(t,\tau)+...
\end{equation}
around the steady flow at $\Rey_c$, namely the base flow $\mathbf{U}_0=(\mathbf{u}_0,p_0)^T$. 
Here the approximate equality signifies the fact that asymptotic expansions are not necessarily convergent over the whole domain (see \cite{vandyke75perturbation}). We assume the forcing to have the form
\begin{equation}\label{eq:forcing}
    \mathbf{f}(t, \tau)=E'(\tau)e^{\imath \omega_f t} \mathbf{f}_E + c.c., \footnote{c.c. stands for complex conjugate}
\end{equation}
where $\omega_f$ is the forcing frequency and $\mathbf{f}_E$ is the complex spatial structure of the forcing.

The complex forcing amplitude $E'(\tau)$ is evolving on the slow timescale $\tau$, and scales as $E'(\tau)=\sqrt{\epsilon}^i E(\tau)$, where $i$ depends on the relation between the forcing frequency $\omega_f$ and the shedding frequency $\omega_0$ of the global mode at $Re_c$. 

Substituting the asymptotic expansions of the state~\eqref{eq:expansion} and the forcing~\eqref{eq:forcing} 
into the governing equation~\eqref{eq:NS} leads to a series
of linear equations at various orders $\sqrt{\epsilon}^i$ for $i=0,1,2,3,...$, which are successively solved. We adopt the scaling approach from~\cite{fauve2009} and \cite{sipp2012open}, where different scalings of $E'(\tau)$ are used for different classes of frequencies to avoid degenerate operator equations at orders $\sqrt{\epsilon}^1$ and $\sqrt{\epsilon}^2$.
In all of the cases, the Stuart-Landau equation follows from compatibility conditions at order $\sqrt{\epsilon}^3$. Depending on the forcing frequency however, different forcing terms are obtained. The constructed Stuart-Landau equations are of the same form as in~\cite{sipp2012open} where $E'$ is held constant, except that here the forcing amplitude $E'(\tau)$ is allowed to be time-varying, which enables the design of closed-loop control laws. We next provide a detailed derivation of the Stuart-Landau equation for resonant forcing frequencies near $\omega_0$ and briefly present the corresponding results for the other frequencies, which are obtained analogously.

\subsection{Forcing near the resonant frequency $ \omega_0$}\label{sec:resonant}%
Following the scaling suggested by~\cite{fauve2009}, the forcing amplitude is 
\begin{equation}\label{eq:scaling_resonant}
E'(\tau):=\sqrt{\epsilon}^3E(\tau)
\end{equation}
when $\omega_f$ is close to the frequency of the global mode $\omega_0$ at $\Rey_c$. In particular, we assume that $\omega_f=\omega_0+\epsilon \Omega$ for some frequency $\Omega$.
Inserting the expansion~\eqref{eq:expansion} and the forcing~\eqref{eq:forcing}, \eqref{eq:scaling_resonant} with $\epsilon$ as given by~\eqref{eq:reynolds} into the Navier-Stokes equations~\eqref{eq:NS_matrix1}, we obtain a sequence of equations at orders $\sqrt{\epsilon}^i$ for $i=0,1,2,3...$. 

\subsubsection{Order $\sqrt{\epsilon}^0$}
At the zeroth order $\sqrt{\epsilon}^0$, we obtain the steady nonlinear Navier-Stokes equation 
\begin{equation}\label{eq:base}
      \mathscr{N}(\mathbf{U}_0, \Rey_c) = \left(
\begin{array}{cc}
  -\nabla \mathbf{u}_0\cdot \mathbf{u}_0-\nabla p_0+ \frac{1}{\Rey_c}\Delta \mathbf{u}_0\\
  \displaystyle
    \nabla \cdot \mathbf{u}_0 \\
\end{array}  \right)=\mathbf{0}
\end{equation}
at the critical Reynolds number $\Rey_c$. This equation is satisfied by default by the base flow $\mathbf{U}_0$, which is an equilibrium at $\Rey_c$.

\subsubsection{Order $\sqrt{\epsilon}$}
At order $\sqrt{\epsilon}$, we obtain the homogeneous linear equation
\begin{equation}\label{eq:NSlin_forced}
\Big(\frac{\partial}{\partial t} \mathscr{E} - \mathscr{A} \Big)\mathbf{U}_1=\mathbf{0}, 
\end{equation}
where
\begin{equation*}
   \mathscr{A}=\left(
\begin{array}{cc}
   -\nabla () \cdot \mathbf{u}_0-\nabla \mathbf{u}_0\cdot ()+ \frac{1}{\Rey_c}\Delta &   \nabla  \\
  \displaystyle
   \nabla^T &  0 \\
\end{array}  \right)
\end{equation*}
denotes the linearized Navier-Stokes operator around the base flow $\mathbf{U}_0$. To expand the general solution of \eqref{eq:NSlin_forced} in terms of eigenvectors $\mathbf{U}$ of the operator pencil 
\begin{equation*}
\mathscr{K}_{\lambda}=\lambda\mathscr{E}-\mathscr{A}
\end{equation*}
where $\lambda=\sigma+\imath \omega$, we solve the generalized eigenvalue problem
\begin{equation}\label{eq:eigenvalue}
\mathscr{K}_{\lambda}\mathbf{U}=0.
\end{equation}

Since we assume that the first Hopf bifurcation occurs at $\Rey_c$, the spectrum
of $\mathscr{K}_{\lambda}$ contains a single marginally stable eigenvalue pair $\pm \imath \omega_0$ with the associated eigenvector $\mathbf{U}_1^A$ (and its complex conjugate) given by 
 \begin{equation*}\label{eq:order1}
     \mathscr{K}_{\imath \omega_0}\mathbf{U}_1^{A}=\mathbf{0},
 \end{equation*}
while the rest of the eigenvalues are stable. For this reason, we retain only the solution along this eigenspace, which governs the long-term dynamics. It
is written as
\begin{equation*}
    \mathbf{U}_{1}(t, \tau) = A(\tau) e^{\imath \omega_0 t}\mathbf{U}_1^A + c.c.,
  \end{equation*}
  where the amplitude $A(\tau)$ evolves on the slow timescale.
The eigenvector $\mathbf{U}_1^{A}$ is termed as the global mode or the first harmonic in the literature.

\subsubsection{Order $\sqrt{\epsilon}^2$}
At order $\sqrt{\epsilon}^2$, we have the following linear inhomogeneous equation
\begin{align*}
       \Big(\frac{\partial}{\partial t} \mathscr{E} - \mathscr{A}\Big)\mathbf{U}_2=\mathbf{F}_2^1+\lvert A (\tau)\rvert^2 \mathbf{F}^{\lvert A \rvert^2} +\big(A(\tau)^2 e^{\imath 2\omega_0t}\mathbf{F}^{A^2}+c.c.\big),
\end{align*}
where the right-hand side terms are
\begin{align*}
    \mathbf{F}^{\lvert A \rvert^2}:=\mathscr{P}\big(
     -\nabla \mathbf{u}_1^A \cdot \overline{\mathbf{u}_1^A}-\nabla \overline{\mathbf{u}_1^A} \cdot \mathbf{u}_1^A \big),~\mathbf{F}^{A^2}:=\mathscr{P}\big(
     -\nabla \mathbf{u}_1^A \cdot \mathbf{u}_1^A  \big),~\mathbf{F}_2^1:=\mathscr{P}\big(
        - \Delta \mathbf{u}_0 \big).
\end{align*}
We can write $\mathbf{U}_2(t, \tau)$ as the superposition of the responses of the linear system to each forcing term, i.e.
\begin{align}
       \mathbf{U}_2(t, \tau)= \mathbf{U}_2^1+\lvert A (\tau) \rvert^2 \mathbf{U}_2^{\lvert A \rvert^2} +A(\tau)^2 e^{\imath 2\omega_0t}\mathbf{U}_2^{A^2}+c.c.,\label{eq:u2_forced}
\end{align}
where the velocity-pressure fields are determined by solving the equations
\begin{align} \label{eq:order2}
    \mathscr{K}_0 \mathbf{U}_2^1= \mathbf{F}_2^1,\quad
    \mathscr{K}_0 \mathbf{U}_2^{\lvert A \rvert^2}=\mathbf{F}^{\lvert A \rvert^2}, \quad
    \mathscr{K}_{\imath 2\omega_0} \mathbf{U}_2^{ A^2}=\mathbf{F}^{A^2}. 
\end{align}
By the conditions of the Hopf bifurcation theorem~\cite[Theorem 8.25]{chicone2006ordinary}, $\lambda=0$ and $\lambda=\imath 2\omega_0$ are not in the spectrum of the operator pencil $\mathscr{K}_{\lambda}$, and therefore these equations always admit a unique solution.

Here $\mathbf{U}_{2}^1$ is the first-order approximation of the modification of the neutral equilibrium $\mathbf{U}_0$ at $\epsilon=0$ to an unstable equilibrium at $\epsilon>0$. The zeroth (mean flow) harmonic $\lvert A (\tau) \rvert^2 \mathbf{U}_2^{\lvert A \rvert^2}$, which evolves on the slow timescale $\tau$, results from the nonlinear interaction between the first harmonic $A (\tau) e^{\imath \omega_0 t}\mathbf{U}_1^A$ with its conjugate. It corresponds to the difference between the mean flow and the base flow (\cite{sipp2007global}). Finally, the second harmonic $A (\tau) ^2 e^{\imath 2\omega_0t}\mathbf{U}_2^{A^2}$ evolving on the slow and fast timescale (with frequency $2\omega_0$)
stems from the nonlinear interaction between the first harmonic with itself.

\subsubsection{Order $\sqrt{\epsilon}^3$}
At order $\sqrt{\epsilon}^3$ we obtain the inhomogeneous linear equation 
\begin{align}\label{eq:thirdorder}
     \Big(\frac{\partial}{\partial t} \mathscr{E}-\mathscr{A} \Big)\mathbf{U}_3 = &-\mathscr{E} \frac{d A}{d \tau} e^{\imath \omega_0 t} \mathbf{U}_1^A+A (\tau) e^{\imath \omega_0 t} \mathbf{F}^A+A (\tau)\lvert A(\tau) \rvert^2 e^{\imath \omega_0 t} \mathbf{F}^{A\lvert A \rvert^2} \nonumber \\ 
     & +E(\tau) e^{\imath \omega_f t} \mathcal{P}\mathbf{f}_E +c.c.+ ...
\end{align}
where the right-hand side terms $\mathbf{F}^A$ and $\mathbf{F}^{A \lvert A \rvert^2}$ are defined as
\begin{align*}
           \mathbf{F}^A & :=\mathscr{P}\big(
          -\nabla \mathbf{u}_1^A \cdot \mathbf{u}_2^1- \nabla \mathbf{u}_2^1 \cdot \mathbf{u}_1^A- \Delta \mathbf{u}_1^A \big), \nonumber \\
         \mathbf{F}^{A \lvert A \rvert^2}& :=\mathscr{P} \big(-\nabla \mathbf{u}_1^A \cdot \mathbf{u}_2^{\lvert A \rvert^2}-\nabla \mathbf{u}_2^{\lvert A \rvert^2}\cdot \mathbf{u}_1^A-\nabla \bar{\mathbf{u}}_1^A \cdot \mathbf{u}_2^{A^2}-\nabla \mathbf{u}_2^{A^2} \cdot \bar{\mathbf{u}}_1^A \big). 
\end{align*}
The term $\mathbf{F}^A$ stems from the viscous diffusion of the first harmonic $Ae^{\imath \omega_0 t}\mathbf{U}_1^{A}$ and its interaction with the base flow modification $\mathbf{U}_2^{1}$. These interactions are responsible for the instability of
the flow (\cite{sipp2007global}). The term $\mathbf{F}^{A\lvert A \rvert^2}$  arises due to the interaction of the first
harmonic $A (\tau)e^{\imath \omega_0 t}\mathbf{U}_1^{A}$ with the zeroth harmonic $\lvert A(\tau) \rvert^2 \mathbf{U}_2^{\lvert A \rvert^2}$ and the second harmonic $A(\tau)^2 e^{\imath 2\omega_0t}\mathbf{U}_2^{A^2}$.
As the right-hand side of the equation~\eqref{eq:thirdorder} involves terms that evolve near the resonant frequency, the solution $ \mathbf{U}_3(t, \tau)$ may grow unbounded in time when these terms are generic. 



\subsubsection{Stuart-Landau Equation}

To ensure that the solution will indeed remain bounded, and hence, retain consistency of the asymptotic expansion, 
the terms in the right-hand side of~\eqref{eq:thirdorder} need to be orthogonal to the adjoint eigenvector $\mathbf{U}_1^{A \star}$ of $\mathscr{K}_{\imath \omega_0}$ at all times. This imposes structural constraints on the evolution of the amplitude of the global mode, which needs to satisfy the Stuart-Landau equation 
\begin{equation}\label{eq:SL_resonance}
    \phantom{.}\frac{d A}{d \tau}=a_0 A - a_1 A \lvert A \rvert^2 +a_2 e^{\imath \epsilon \Omega t}E, 
\end{equation}
%
%
where the coefficients $a_0$, $a_1$, and $a_2$ are determined by the orthogonality conditions 
\begin{subequations}
\label{eq:SLcoefficients}
\begin{eqnarray}
   a_0&:=&\frac{\langle \mathbf{U}_1^{A \star},\mathbf{F}^A  \rangle}{\langle \mathbf{U}_1^{A \star},  \mathscr{E} \mathbf{U}_1^A \rangle} \label{eq:a0} \\
     a_1&:=&-\frac{\langle \mathbf{U}_1^{A \star},\mathbf{F}^{A \lvert A \rvert^2 } \rangle}{\langle \mathbf{U}_1^{A \star},  \mathscr{E} \mathbf{U}_1^A \rangle} \label{eq:a1} \\
a_2&:=&\frac{\langle \mathbf{U}_1^{A \star},\mathscr{P}\mathbf{f}_E \rangle}{\langle \mathbf{U}_1^{A \star},  \mathscr{E} \mathbf{U}_1^A \rangle}.\label{eq:a2}
\end{eqnarray}
\end{subequations}
Here, the scalar product of two terms $\mathbf{U}_{\alpha}$ and $\mathbf{U}_{\beta}$ is defined as
\begin{equation}\label{eq:scalar_product}
    \langle \mathbf{U}_{\alpha}, \mathbf{U}_{\beta} \rangle := \iint_{\Omega} (\bar{u}_{\alpha}u_{\beta}+\bar{v}_{\alpha}v_{\beta}+\bar{p}_{\alpha}p_{\beta}) dx dy.
\end{equation}
Thus, the solution of the flow at this order takes the form 
\begin{equation*}
    \mathbf{U}_3(t, \tau)=A(\tau)e^{\imath \omega_0 t}\mathbf{U}_3^A+A(\tau)\lvert A(\tau) \rvert^2 e^{\imath \omega_0 t}\mathbf{U}_3^{A\lvert A \rvert^2}+E(\tau)e^{\imath \omega_f t}\mathbf{U}_3^E+c.c.+ ...
\end{equation*}
where we only write the terms that evolve near the resonant frequency.

\subsubsection{Reduced-order Model}

Above we derived the forced Stuart-Landau equation~\eqref{eq:SL_resonance} for resonant forcing frequencies $\omega_f \approx \omega_0$, which can be rewritten with respect to the fast timescale $t$ as 
\begin{equation}\label{eq:SL_resonance_t}
   \frac{d A}{d t}=\epsilon a_0 A - \epsilon a_1 A \lvert A \rvert^2 + \epsilon a_2e^{\imath \epsilon \Omega t} E. 
   \end{equation}
In real coordinates $\mathbf{X}=[\Re(A), \Im(A)]^T$ with input $\mathbf{q}=[\Re(E), \Im(E)]^T$ this 
takes the form~\eqref{eq:rom} of a surrogate model to the incompressible Navier-Stokes equations~\eqref{eq:NS_matrix}.
The presented weakly nonlinear analysis approximates the solution $\mathbf{U}$ 
as 
\begin{align}
  \mathbf{U}\approx \mathbf{U}_0&+\sqrt{\epsilon} \big(Ae^{\imath \omega_0t}\mathbf{U}_1^A+c.c.\big)+\epsilon \big(\mathbf{U}_2^1+ \lvert A \rvert^2 \mathbf{U}_2^{\lvert A \rvert^2} + \big(A^2 e^{\imath 2\omega_0t}\mathbf{U}_2^{A^2}+ c.c.\big)\big) \nonumber \\
 & + \sqrt{\epsilon}^{3}\big(A e^{\imath \omega_0 t}\mathbf{U}_3^A+A\lvert A \rvert^2 e^{\imath \omega_0 t}\mathbf{U}_3^{A\lvert A \rvert^2} + Ee^{\imath (\omega_0+\epsilon \Omega) t}\mathbf{U}_3^E+c.c.\big)+ ....\label{eq:solution_wna}
\end{align}
From~\eqref{eq:forcing}, \eqref{eq:scaling_resonant}, and~\eqref{eq:solution_wna} we obtain mappings   $\mathbf{q}\mapsto\mathscr{Q}(\mathbf{q})=\mathbf{f}$ and $\mathbf{X}\mapsto \mathscr{G}(\mathbf{X})\approx\mathbf U$ from the input and state space
of the surrogate model to the corresponding spaces of the original system~\eqref{eq:NS_matrix1}. We denote the approximation error by 
\begin{equation}\label{eq:error1}
    \mathbf e\equiv(\mathbf{e}_{\mathbf{u}}, e_p):=\mathbf U-\mathscr{G}(\mathbf{X})
\end{equation}

\subsection{Other Resonant Forcing Frequencies}
Here we discuss other cases of resonant forcing frequencies, which include $\omega_f=0$, $\omega_f \approx 2\omega_0$, and $\omega_f \approx \frac{\omega_0}{2}$. Each of these frequencies requires a different scaling to obtain a forcing term in the Stuart-Landau equation at order $\sqrt{\epsilon}^3$ while avoiding degenerate operators at lower orders.

\subsubsection{Forcing at $\omega_f=0$}
The forcing amplitude at this frequency scales as
\begin{equation*}
    E'(\tau):= \epsilon E (\tau).
\end{equation*}
Solving the corresponding equations at orders $\sqrt{\epsilon}^0$, $\sqrt{\epsilon}^1$, $\sqrt{\epsilon}^2$ and $\sqrt{\epsilon}^3$, $\mathbf{U}$ is obtained in the form 
\begin{align}\label{eq:wna_zero}
    \mathbf{U} \approx\; & \mathbf{U}_0+ \sqrt{\epsilon} \big(A e^{\imath \omega_0 t}\mathbf{U}_1^A + c.c.\big)
\nonumber \\ 
 & +\epsilon \big( \mathbf{U}_2^1+\lvert A \rvert^2 \mathbf{U}_2^{\lvert A \rvert^2} + \big(A^2 e^{\imath 2\omega_0t}\mathbf{U}_2^{A^2}+E \mathbf{U}_2^E + c.c.\big) \big)\nonumber \\
     & +\sqrt{\epsilon}^{3} \big(A e^{\imath \omega_0 t} \mathbf{U}_3^A+A\lvert A \rvert^2 e^{\imath \omega_0 t} \mathbf{U}_3^{A\lvert A \rvert^2}+AE e^{\imath \omega_0 t} \mathbf{U}_3^{AE} +
     A\bar{E} e^{\imath \omega_0 t} \mathbf{U}_3^{A\bar{E}} + c.c.\big) + ..., 
\end{align} 
where $\mathbf{U}_0$, $\mathbf{U}_1^A$, $\mathbf{U}_2^1$, $\mathbf{U}_2^{\lvert A \rvert^2}$, $\mathbf{U}_2^{A^2}$ are the same as in Section~\ref{sec:resonant}, while the
forcing response $\mathbf{U}_2^E$ is obtained via
\begin{align*}
\mathscr{K}_{0}\mathbf{U}_2^{E}&=\mathscr{P}  \mathbf{f}_E.
\end{align*} 
At order $\sqrt{\epsilon}^3$, the terms $\mathbf{U}_3^A$, $\mathbf{U}_3^{A\lvert A \rvert^2}$, $\mathbf{U}_3^{A E}$ and $\mathbf{U}_3^{A \bar{E}}$ need to satisfy again compatibility conditions  which yield the Stuart-Landau equation
\begin{equation}\label{eq:SL_omega0}
    \frac{d A}{d t}=\epsilon(a_0-a_2E-a_3\bar{E})A- \epsilon a_1 A \lvert A \rvert^2.
    \end{equation}
As the coefficients $a_0$ and $a_1$ depend only on the unforced modes, they are independent of the forcing and are given as in Section~\ref{sec:resonant} by~\eqref{eq:a0} and~\eqref{eq:a1}, while 
\begin{align*}
 a_2&:=\frac{\langle \mathbf{U}_1^{A \star}, \mathscr{P}\big( \nabla \mathbf{u}_1^A \cdot \mathbf{u}_1^E+\nabla \mathbf{u}_1^E\cdot \mathbf{u}_1^A 
     \big) \rangle}{\langle \mathbf{U}_1^{A \star},  \mathscr{E} \mathbf{U}_1^A \rangle} \\
      a_3&:=\frac{\langle \mathbf{U}_1^{A \star},\mathscr{P} \big( \nabla \mathbf{u}_1^A \cdot \overline{\mathbf{u}_1^E}+\nabla \overline{\mathbf{u}_1^E}\cdot \mathbf{u}_1^A 
     \big) \rangle}{\langle \mathbf{U}_1^{A \star},  \mathscr{E} \mathbf{U}_1^A \rangle}.
\end{align*}
From~\eqref{eq:SL_omega0}, we can see that for $\omega_f=0$, the forcing term appears bilinearly 
in the dynamics of the global mode, in contrast to its purely additive structure in~\eqref{eq:SL_resonance_t}.  

\subsubsection{Forcing near $2\omega_0$}
For a forcing frequency near $2\omega_0$, in particular, when $ \omega_f = 2\omega_0+\epsilon \Omega$ for some frequency $\Omega$, we have
\begin{equation*}
    E'(\tau):= \epsilon E (\tau)
\end{equation*}
and $\mathbf{U}$ is of the form  
\begin{align}\label{eq:wna_double}
    \mathbf{U} \approx ~\mathbf{U}_0&+\sqrt{\epsilon} \big(A e^{\imath \omega_0 t}\mathbf{U}_1^A + c.c.\big)
\nonumber \\ 
 & +\epsilon \big(\mathbf{U}_2^1+\lvert A \rvert^2 \mathbf{U}_2^{\lvert A \rvert^2} + \big(A^2 e^{\imath 2\omega_0t}\mathbf{U}_2^{A^2} +E e^{\imath (2\omega_0+\epsilon \Omega)t}\mathbf{U}_2^E + c.c.\big )\big) \nonumber \\
     & +\sqrt{\epsilon}^{3} \big( A e^{\imath \omega_0 t} \mathbf{U}_3^A+A\lvert A \rvert^2 e^{\imath \omega_0 t} \mathbf{U}_3^{A\lvert A \rvert^2}+
     \bar{A}E e^{\imath (\omega_0+\epsilon \Omega) t} \mathbf{U}_3^{\bar{A}E} + c.c. \big) + .... 
\end{align} 

Here, the forcing response is determined by the solution of
\begin{align*}
\mathscr{K}_{\imath (2\omega_0+\epsilon \Omega)}\mathbf{U}_2^{E}&=\mathscr{P}  \mathbf{f}_E.
\end{align*} 
The Stuart-Landau equation for this forcing frequency case is 
\begin{equation*}
  \frac{d A}{d t}=\epsilon a_0 A- \epsilon a_1  A \lvert A \rvert^2 -\epsilon a_2 e^{\imath \epsilon \Omega t}\bar{A}E
\end{equation*}
with the coefficient 
\begin{align*}
 a_2:=\frac{\langle \mathbf{U}_1^{A \star}, \mathscr{P}\big( \nabla \overline{\mathbf{u}_1^A} \cdot \mathbf{u}_1^E+ \nabla \mathbf{u}_1^E \cdot \overline{\mathbf{u}_1^A} 
     \big) \rangle}{\langle \mathbf{U}_1^{A \star},  \mathscr{E} \mathbf{U}_1^A \rangle}
\end{align*}
determined by the interaction of the
conjugate of the global mode with the forcing response.

\subsubsection{Forcing near $\frac{\omega_0}{2}$}
The last resonant frequency case concerns forcing near $\frac{\omega_0}{2}$. Again, we assume that $\omega_f=\frac{\omega_0}{2}+\epsilon \Omega$. The appropriate scaling of $E'(\tau)$ is
\begin{equation*}
    E'(\tau):= \epsilon^{3/4}  E(\tau)
\end{equation*}
and $\mathbf{U}$ is of the form 
\begin{align}\label{eq:wna_half}
    \mathbf{U} \approx ~\mathbf{U}_0&+\sqrt{\epsilon}\big(A e^{\imath \omega_0 t}\mathbf{U}_1^A + c.c.\big)
\nonumber \\ 
& +\epsilon^{3/4} \big(E e^{\imath (\omega_0/2+\epsilon \Omega) t} \mathbf{U}_2^E + c.c. \big) \nonumber \\
 & +\epsilon \big(\mathbf{U}_2^1+\lvert A \rvert^2 \mathbf{U}_2^{\lvert A \rvert^2} +A^2 e^{\imath 2\omega_0t}\mathbf{U}_2^{A^2} \big)\nonumber \\
     & +\sqrt{\epsilon}^{3} \big(A e^{\imath \omega_0 t} \mathbf{U}_3^A+A\lvert A\rvert^2 e^{\imath \omega_0 t} \mathbf{U}_3^{A\lvert A \rvert^2}+
     E^2 e^{\imath (\omega_0+2 \epsilon \Omega) t} \mathbf{U}_3^{E^2} + c.c.\big) + ..., 
\end{align} 
where the forcing response is obtained from
\begin{align*}
\mathscr{K}_{\imath (\omega_0/2+ \epsilon \Omega)}\mathbf{U}_2^{E}&=\mathscr{P}  \mathbf{f}_E. 
\end{align*} 
The Stuart-Landau equation is given as
\begin{equation}\label{eq:SL_omegahalf}
  \frac{d A}{d t}=\epsilon a_0 A- \epsilon  a_1 A \lvert A \rvert^2 - \epsilon a_2 e^{\imath 2 \epsilon \Omega}E^2,
\end{equation}
where the coefficient
\begin{align*}
 a_2:=\frac{\langle \mathbf{U}_1^{A \star}, \mathscr{P}\big(\nabla \mathbf{u}_1^E \cdot \mathbf{u}_1^E
     \big) \rangle}{\langle \mathbf{U}_1^{A \star},  \mathscr{E} \mathbf{U}_1^A \rangle} 
\end{align*}
stems from the nonlinear interaction of
the forcing response with itself. Here the forcing enters the Stuart-Landau equation~\eqref{eq:SL_omegahalf} as a nonlinear additive term, in analogy to ~\eqref{eq:SL_resonance_t} where the forcing is close to the natural frequency of the flow.
\subsection{Non-Resonant Forcing}
For a non-resonant forcing, where 
$\omega_f\ne 0$ or $\omega_f \not\approx\omega_0 ,\frac{\omega_0}{2},2\omega_0$,  
the forcing amplitude scales as
\begin{equation}\label{eq:scaling_nonresonant}
    E'(\tau):=\sqrt{\epsilon}E(\tau).
\end{equation}
The equations at orders $\sqrt{\epsilon}^0$, $\sqrt{\epsilon}$, $\sqrt{\epsilon}^2$ and $\sqrt{\epsilon}^3$ are successively solved as in the resonant cases, and $\mathbf{U}$ is obtained in the form  
\begin{align}\label{eq:wna_nonres}
    \mathbf{U} \approx ~\mathbf{U}_0&+ \sqrt{\epsilon} \big(A e^{\imath \omega_0 t}\mathbf{U}_1^A+E e^{\imath \omega_f t}\mathbf{U}_1^E + c.c.\big) 
\nonumber \\ 
     & +\epsilon \big(\mathbf{U}_2^1+\lvert A \rvert^2 \mathbf{U}_2^{\lvert A \rvert^2} + \big(A^2 e^{\imath 2\omega_0t}\mathbf{U}_2^{A^2}+ c.c.\big) \nonumber  \\
&+\lvert E \rvert^2 \mathbf{U}_2^{\lvert E \rvert^2} +\big(E^2 e^{\imath 2\omega_f t}\mathbf{U}_2^{E^2}+c.c.\big) \nonumber  \\
       &+\big(AEe^{\imath (\omega_0+\omega_f)t}\mathbf{U}_2^{AE}+ A\bar{E}e^{\imath (\omega_0-\omega_f)t}\mathbf{U}_2^{A\bar{E}}+c.c.\big) \big) \nonumber \\
     & +\epsilon^{3/2} \big( A e^{\imath \omega_0 t} \mathbf{U}_3^A+A\lvert A \rvert^2 e^{\imath \omega_0 t} \mathbf{U}_3^{A\lvert A \rvert^2}
    +A\lvert E\rvert^2 e^{\imath \omega_0 t} \mathbf{U}_3^{A \lvert E\rvert^2}+ c.c.\big)+ ... 
\end{align} 
where the velocity-pressure fields $\mathbf{U}_1^E$, $\mathbf{U}_2^{\lvert E \rvert^2}$, $\mathbf{U}_2^{E^2}$, $\mathbf{U}_2^{AE}$, $\mathbf{U}_2^{A\bar{E}}$ are obtained by solving the following operator equations
\begin{align*}
\mathscr{K}_{\imath \omega_f}\mathbf{U}_1^{E}&=\mathscr{P}  \mathbf{f}_E, \\
 \mathscr{K}_{0}\mathbf{U}_2^{\lvert E \rvert^2} &=   \mathscr{P}\big(
         -\nabla \mathbf{u}_1^E \cdot \overline{\mathbf{u}_1^E}-\nabla \overline{\mathbf{u}_1^E} \cdot \mathbf{u}_1^E \big),\\
      \mathscr{K}_{2\imath \omega_f}\mathbf{U}_2^{E^2} &=   \mathscr{P}\big(
         -\nabla \mathbf{u}_1^E \cdot \mathbf{u}_1^E \big),   \\
    \mathscr{K}_{\imath (\omega_0+\omega_f)}\mathbf{U}_2^{AE} &= \mathscr{P}\big( -\nabla \mathbf{u}_1^A \cdot \mathbf{u}_1^E-\nabla \mathbf{u}_1^E \cdot \mathbf{u}_1^A 
     \big), \\
    \mathscr{K}_{\imath (\omega_0-\omega_f)}\mathbf{U}_2^{A\bar{E}} &= \mathscr{P}\big( -\nabla \mathbf{u}_1^A \cdot \overline{\mathbf{u}_1^E}-\nabla \overline{\mathbf{u}_1^E}\cdot \mathbf{u}_1^A 
     \big).
\end{align*} 
As in the previous cases, at order $\sqrt{\epsilon}^3$, the terms $\mathbf{U}_3^A$, $\mathbf{U}_3^{A\lvert A \rvert^2}$ and $\mathbf{U}_3^{A \lvert E\rvert^2}$ need to satisfy compatibility conditions which yield the Stuart-Landau equation 
\begin{equation*}
    \frac{d A}{d t}=\epsilon (a_0- a_2 \lvert E \rvert^2) A - \epsilon  a_1 A \lvert A\rvert^2 
\end{equation*}
with 
\begin{equation*}
   a_2:=\frac{\big\langle \mathbf{U}_1^{A \star},\mathscr{P}\big(\nabla \mathbf{u}_1^A \cdot \mathbf{u}_2^{ \lvert E\rvert^2}+\nabla \mathbf{u}_2^{ \lvert E\rvert^2} \cdot \mathbf{u}_1^A  +\nabla \mathbf{u}_1^E \cdot \mathbf{u}_2^{A \bar{E}}+\nabla \mathbf{u}_2^{A \bar{E}} \cdot \mathbf{u}_1^E \big) \big\rangle}{\langle \mathbf{U}_1^{A \star},  \mathscr{E} \mathbf{U}_1^A  \rangle}.      
\end{equation*}
As in the case of zero-frequency forcing and forcing near $2 \omega_0$, the input acts multiplicatively on the linear part of the dynamics. 




\section{Closed-Loop Control}\label{sec:control}
Our control objective is to stabilize the flow described by~\eqref{eq:NS_matrix} by steering the velocity-pressure field to its steady state $\mathbf{U}_b$. As suggested by the approximation of $\mathbf{U}$ 
given by either of~\eqref{eq:solution_wna}, ~\eqref{eq:wna_zero},~\eqref{eq:wna_double},~\eqref{eq:wna_half}, and~\eqref{eq:wna_nonres}, both $A$ and $E$ need to be zero at the equilibrium $\mathbf{U}_b\approx \mathbf{U}_0+\epsilon\mathbf{U}_2^1$.

 \subsection{Stabilization Capabilities across Different Frequencies}
 \label{sec:forcing:comparison}

In the case of zero-frequency forcing, forcing near $2 \omega_0$, and non-resonant frequency, the forcing amplitude cannot converge to zero
while maintaining stability of the Stuart-Landau model as it acts multiplicatively on the linear part of its dynamics.
This implies that although we stabilize the Stuart-Landau model, and therefore the global mode and all the other unsteady terms in~\eqref{eq:wna_zero}, ~\eqref{eq:wna_double},~\eqref{eq:wna_nonres} that arise from its interaction with itself or the forcing response, the flow remains unsteady due to the persisting terms that depend exclusively on the forcing.

For the other two cases, $\omega_f \approx \omega_0$ and $\omega_f \approx \frac{\omega_0}{2}$,
amplitude $E$ enters the Stuart-Landau equation as an additive term which allows bringing $E$ to zero while stabilizing the model. Once both $A$ and $E$ converge to zero, the flow is at its equilibrium $\mathbf{U}_b\approx \mathbf{U}_0+\epsilon\mathbf{U}_2^1$ according to~\eqref{eq:solution_wna} and~\eqref{eq:wna_half}. Therefore, the unsteadiness of the flow is suppressed.

 We choose forcing near the resonant frequency $\omega_f \approx \omega_0$ since this case facilitates characterizing the structure $f_E$ which optimizes the control efficiency, 
as shown in Section~\ref{sec:optimalforcing}. Finding such a forcing structure for the forcing frequency $\omega_f \approx \frac{\omega_0}{2}$ is a nontrivial task and we leave it for future research.

\subsection{Closed-loop Control Design}
We design an output-feedback controller for our full-order model in three successive steps. First, we estimate the reduced state from the velocity measurements. Then we feed this estimate into the controller, which determines the control amplitude $E$ of  
the surrogate model \eqref{eq:SL_resonance_t}. 
The last step is to refine this control input into the forcing $f$ of the full-order model via~\eqref{eq:forcing} and \eqref{eq:scaling_resonant}.

In particular, we design a model predictive controller, which seeks to optimally steer the amplitude of the global mode $A$ to zero while incorporating suitable constraints for the forcing amplitude $E$. The latter include soft constraints on the magnitude of the forcing and its rate of change, which are introduced as penalty terms in the objective function of the optimization problem. This enables us to to bring the control input $E$ to zero while naturally complying with our modelling requirement of small temporal gradients of $E$, to be consistent with the separation of timescales assumed in the derivation of the reduced-order model. 

\subsubsection{Model Predictive Control}\label{sec:mpc}

To determine the inputs of the model predictive controller, we consider a discretized version of the forced Stuart-Landau model~\eqref{eq:SL_resonance} with sampling period $\Delta t$ and zero-order hold of the complex input amplitude $E$. We assume that $\Delta t$ is sufficiently large compared to the fast timescale,  
 so that potential jumps of $E$ across the sampling times have no significant impact on the validity of the weakly nonlinear analysis. 

MPC relies on recursively solving a finite-horizon optimal control problem at each time instant and keeping only the first input of the resulting optimization problem. Considering a horizon of $m$ time steps and using the notation $\mathbf{X}_j$ and $\mathbf{q}_j$ for the state and input of the time-discretized Stuart-Landau model at time $t_j=j\Delta t$, we seek to minimize the quadratic cost function
\begin{equation}\label{eq:costfunction}
    J(\mathbf{X}_{[j:j+m]},\mathbf{q}_{[j:j+m-1]}):= \sum_{k=1}^m\| \mathbf{X}_{j+k}-\mathbf{X}^{\ast} \|^2_{\mathbf{Q}}+ \sum_{k=0}^{m-1}\big( \| \mathbf{q}_{j+k}\|^2_{\mathbf{R_u}}+ \| \Delta \mathbf{q}_{j+k}\|^2_{\mathbf{R_{\Delta u}}}\big)
\end{equation}
subject to the dynamics of the system. Here $\mathbf{q}_{[j:j+m-1]}:=(\mathbf{q}_j,..., \mathbf{q}_{j+m-1})$ denotes the sequence of control actions applied over the horizon and $\mathbf{X}_{[j:j+m]}:=(\mathbf{X}_j,...,\mathbf{X}_{j+m})$ denotes the corresponding state sequence of the system. 

Since the model predictive controller is used for the output feedback of the full-order plant, its initial state satisfies the constraint $\mathbf{X}_j=\widetilde{\mathbf{X}}_j$, where the reduced state $\widetilde{\mathbf{X}}_j$ 
is estimated from the measurement 
$\mathbf M(\mathbf U(t_j))$
of the flow field (cf. \eqref{eq:NS_matrix2}) at time $t_j$.
The future states $\mathbf{X}_{j+1},...,\mathbf{X}_{j+m}$ are predicted by the time-discretized version of the Stuart-Landau model~\eqref{eq:SL_resonance} and the inputs are selected in such a way that they minimize the corresponding cost \eqref{eq:costfunction}. Then the first input $\mathbf q_j$ is applied to the system and the same process is repeated recursively. Each term in~\eqref{eq:costfunction} 
is the weighted semi-norm of a vector $\mathbf{x}$, denoted by $\| \mathbf{x} \|^2_{\mathbf{S}} :=\mathbf{x}^T\mathbf{S}\mathbf{x}$ for some positive semi-definite matrix $\mathbf{S}$.
The cost function $J$ 
penalizes the deviations of the
predicted state $\mathbf{X}_{j+k}$ from the reference state $\mathbf{X}^{\ast}= [0,0]$ as well as the magnitude 
of the input $\mathbf{q}_{j+k}$ and its increments $\Delta \mathbf{q}_{j+k}=\mathbf{q}_{j+k}-\mathbf{q}_{j+k-1}$, with the first computed using the previously selected input $\mathbf{q}_{j-1}$ of the MPC recursion. 
The corresponding weight matrix $\mathbf{Q}$ is positive definite, while $\mathbf{R_{u}}$ and $\mathbf{R_{\Delta u}}$ are positive semi-definite.

\begin{figure}
\centering
\includegraphics[width=\textwidth, trim = 1cm 15cm 1cm 0cm, clip=true]{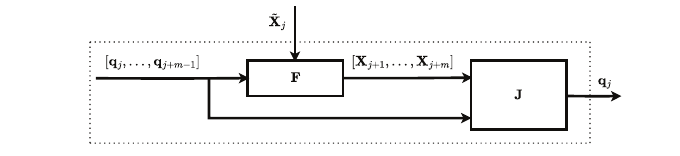}
\caption{Schematics of the model-predictive controller. At each time step $t_j$, the future states $\mathbf{X}_{j+1},...,\mathbf{X}_{j+m}$ are predicted by the surrogate model $\mathbf{F}$ 
with the initial state $\mathbf{X}_j$ set equal to its estimate $\widetilde{\mathbf{X}}_j$ from the flow measurements. The inputs are determined by minimizing the cost function $J$. The first input $\mathbf{q}_j$ is applied to the
system and the same process is repeated at the next time step.}
\end{figure}

\subsubsection{Global mode amplitude estimation from measurements}

Here we discuss the estimation $\tilde{A}$ of the global mode amplitude $A$ from the flow measurements $\mathbf{M}$. Since the dynamics of the reduced order model are only two-dimensional, it is typically not required to determine its state using a dynamic estimator, such as an observer or a Kalman filter. 
There is however a discrepancy between the state estimated from measurements and the one given by the reduced-order model,  
due to the approximation error $\mathbf{e}$ of the model, which is given by~\eqref{eq:error1}. 


We consider the ideal case where we measure the full velocity field and the case where we only measure the velocity at a finite number of points.
%
In both cases, we consider the map 
\begin{align}\label{eq:Gmap}
\mathscr G(\mathbf X):=\mathbf{U}_0+\sqrt{\epsilon}\big(Ae^{\imath \omega_0t}\mathbf{U}_1^A+c.c.\big) +\epsilon \mathbf{U}_2^1
\end{align}
from the surrogate to the full-order model, which retains the terms up to second order from~\eqref{eq:solution_wna} that are linear in $A$. 
%
We also consider a dual map $\mathscr{S}$ from the measurement space of the full-order model to the state space of the surrogate model, which provides the estimation of the global mode amplitude from the velocity measurements.

If the whole velocity field is known, i.e. $\mathbf{M}=(\mathbf{u},0)^T$, then 
\begin{equation}\label{eq:Atilde}
    \mathscr{S}(\mathbf{M}) \equiv  \tilde{A}:= \frac{1}{\sqrt{\epsilon}}\langle \mathbf{u}_1^{A \star}, \mathbf{u}-\mathbf{u}_0- \epsilon \mathbf{u}_2^1\rangle e^{-\imath \omega_0 t},
\end{equation}
where the adjoint velocity mode $\mathbf{u}_1^{A \star}$ is normalized, namely, $\langle \mathbf{u}_1^{A \star}, \mathbf{u}_1^A  \rangle=1$.  
%
%
If we consider instead a finite number of point velocity measurements $\mathbf{M}=\{(\mathbf{u}(\mathbf{x}_i),0)^T\}_{i=1}^n$,  then
\begin{align}
\mathscr{S}(\mathbf{M}) & \equiv \begin{bmatrix}
 \Re (\tilde{{A}}_p) \\
    \Im (\tilde{{A}}_p)
\end{bmatrix} \nonumber \\
& :=
\frac{1}{2 \sqrt{\epsilon}} e^{-\imath \omega_0t} \begin{bmatrix}
    \Re (\mathbf{u}_1^A(\mathbf{x}_1)) & -\Im (\mathbf{u}_1^A(\mathbf{x}_1)) \\
    \vdots & \vdots \\
        \Re (\mathbf{u}_1^A(\mathbf{x}_n)) & -\Im (\mathbf{u}_1^A(\mathbf{x}_n))
\end{bmatrix} ^{\dagger}\begin{bmatrix}
\mathbf{u}(\mathbf{x}_1)- \mathbf{u}_0(\mathbf{x}_1)-\epsilon \mathbf{u}_2^1(\mathbf{x}_1) \\
 \vdots \\
  \mathbf{u}(\mathbf{x}_n)- \mathbf{u}_0(\mathbf{x}_n)-\epsilon \mathbf{u}_2^1(\mathbf{x}_n) \\
\end{bmatrix}. \label{eq:measurements}
\end{align}

\subsection{Optimal Forcing Structure}\label{sec:optimalforcing}
The choice of the forcing structure $\mathbf{f}_E$ directly affects control efficiency through the coefficient $a_2$. In particular, due to the structure of the forced Stuart-Landau dynamics, which are fully actuated with respect to $E$, the higher the magnitude $|a_2|$, the lower the magnitude of the forcing amplitude $|E|$ required to bring $|A|$ to zero. To make this argument precise, let $t\mapsto E_\alpha(t)$ be a (potentially optimal) control trajectory that stabilizes the system, which is subject to the forcing structure $\mathbf f_{E,\alpha}$, and hence, has a forcing coefficient  $a_{2,\alpha}$. When the system is subject to another forcing structure $\mathbf f_{E,\beta}$ with a smaller coefficient $a_{2,\beta}$, the input $t\mapsto  E_\beta(t)= a_{2,\beta}/a_{2,\alpha}E_\alpha(t)$ will have a smaller magnitude and yield the exact same state trajectory. This means that we can always achieve the same control objective as in the first case with a smaller control effort. This is also consistent with the MPC cost in \eqref{eq:costfunction}, which will be respectively lower when the weight matrices $\mathbf{R_{u}}$ and $\mathbf{R_{\Delta u}}$ are multiples of the identity matrix.

We thus seek a forcing structure $\mathbf{f}_E$ of unit energy $\langle \mathbf{f}_E, \mathbf{f}_E \rangle=1$ that gives the highest magnitude of $|a_2|$.
For the resonant case where $\omega_f \approx \omega_0$, the highest $|a_2|$ is obtained when the forcing structure is in the same direction as the velocity component of the adjoint global mode $\mathbf{u}_1^{A \star}$. This follows from the definition~\eqref{eq:a2} of the $a_2$ coefficient. Thus, we obtain an optimal forcing structure $\mathbf{f}_E$ by normalizing $\mathbf{u}_1^{A \star}$, i.e., by selecting
\begin{equation}\label{eq:optimal_forcing}
 \mathbf{f}_E:=\frac{\mathbf{u}_1^{A \star}} {\sqrt{\langle \mathbf{u}_1^{A \star}, \mathbf{u}_1^{A \star} \rangle}}.   
\end{equation}
This choice is unique up to a constant from the complex circle.


\section{Numerical Setup}\label{sec:numericalsetup}

In this section, we present the numerical setup for calculating the coefficients of the Stuart-Landau model~\eqref{eq:SL_resonance} of the flow over a cylinder and performing the direct numerical simulation (DNS) of the incompressible Navier-Stokes equations~\eqref{eq:NS_matrix1} for its validation.

\subsection{Computational Domain and Mesh}\label{sec:computationaldomain}

\begin{figure}
\centering
\includegraphics[width=\textwidth, trim = 1cm 13.8cm 1cm -0.3cm, clip=true]{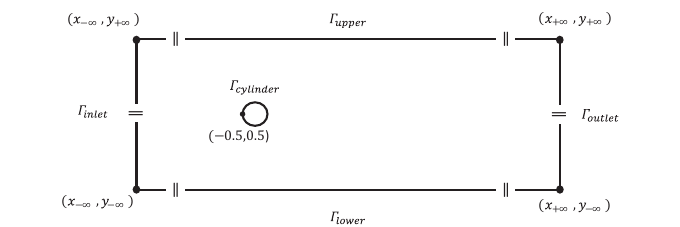}
\caption{Computational domain.
The streamwise and transverse coordinates 
$x_{-\infty}$/$x_{+\infty}$, and $y_{-\infty}$/$y_{+\infty}$, determine respectively the location of the inlet/outlet and lower/upper boundaries. The primary flow direction is from left to right.
}\label{fig:mesh}
\end{figure}

The computational domain is shown in Figure~\ref{fig:mesh} where the origin of the Cartesian coordinate system $\mathbf{x}=(x,y)$ is placed at the center of the cylinder with diameter $D=1$. The domain extends to $x_{-\infty}=-60$ in the upstream direction, $x_{+\infty}=200$ in the downstream direction and from $y_{-\infty}=-30$ to $y_{+\infty}=30$ in the transverse direction. These spatial parameters are adopted from~\cite{sipp2007global} where a detailed analysis of the influence of the computational domain on the parameters of weakly nonlinear analysis was carried out. We use Dirichlet boundary conditions $\mathbf{u}=(1,0)$ on the inlet $\Gamma_{\rm inlet}$, no-slip boundary conditions $\mathbf{u}=(0,0)$ on the cylinder boundary $\Gamma_{\rm cylinder}$, the standard freestream boundary condition $(p-Re^{-1}\frac{\partial{u}}{\partial{x}}=0, \frac{\partial{v}}{\partial{x}}=0)$ at the outlet $\Gamma_{\rm outlet}$ and the symmetric boundary conditions $(\frac{\partial{u}}{\partial{y}}=0, v=0)$ on the upper and lower boundary, $\Gamma_{\rm upper}$ and $\Gamma_{\rm lower}$.

The Navier-Stokes equations are spatially discretized with the Finite Element Method (FEM) using the FreeFEM++ software (see www.freefem.org). We choose Taylor–Hood elements P2 for velocity and P1 for the pressure. The mesh is unstructured and the gridpoint distribution is determined by the automated mesh adaptation method AdaptMesh in FreeFEM++, which relies on the Delaunay-Voronoi algorithm. Within this algorithm, the gridpoint distribution is determined by using a metric matrix. To optimize the mesh we adapt its resolution to the base flow and the structure of the direct and adjoint global modes. The mesh has a total number of $N_t=114382$ triangles and $N_{\rm dof}=516134$ degrees of freedom, which makes it significantly lighter than the mesh used in~\cite{sipp2007global}.
To invert the $N_{\rm dof}\times N_{\rm dof}$ matrices of the discretized operators in the weakly nonlinear analysis, we use the  library, which relies on a sparse direct LU solver (see~\cite{davis2004algorithm}).  

\subsection{Direct Numerical Simulation}\label{sec:DNS}

The direct numerical simulation of the unsteady incompressible Navier-Stokes equations~\eqref{eq:NS_matrix1} marches the velocity-pressure field $\mathbf{U}$ in time. 
To improve numerical stability, we consider the perturbation form  
\begin{equation}\label{eq:NS_perturbation}
   \mathscr{E} \frac{\partial{\mathbf{U}'}}{\partial t}=
   \mathscr{A} \mathbf{U}' -\mathcal{P}\big(
 \nabla \mathbf{u}'\cdot \mathbf{u}'-\mathbf{f}  \big)
\end{equation}
where $\mathbf{U}'=(\mathbf{u}', p')=\mathbf{U}-\mathbf{U}_b$ contains the velocity and pressure components of the perturbation around the steady solution $\mathbf{U}_b$ (base flow) at some $\Rey$. 
For the time discretization of~\eqref{eq:NS_perturbation}, we use the second-order semi-implicit backward-finite-difference scheme, which gives at time $t_{i+1}=(i+1) \Delta t$ the field 
\begin{equation}\label{eq:NS_perturbation_td}
\begin{aligned}
 \mathbf{U}'(t_{i+1})=&\Big(\frac{3}{2 \Delta t} \mathscr{E}-\mathscr{A}\Big)^{-1} \Big(\frac{2}{\Delta t} \mathscr{E}\mathbf{U}'(t_i)-\frac{1}{2\Delta t} \mathscr{E}\mathbf{U}'(t_{i-1})-
 \\
 &\big.\mathcal{P}\big( 2\nabla \mathbf{u}'(t_i)\cdot \mathbf{u}'(t_i)-\nabla \mathbf{u}'(t_{i-1})\cdot \mathbf{u}'(t_{i-1})-2\mathbf{f}(t_i)+\mathbf{f}(t_{i-1})
   \big)\Big).
\end{aligned}
\end{equation}
We use the spatial discretization and finite elements from Section~\ref{sec:computationaldomain} and the time step $\Delta t=0.05$.

\section{Results}\label{sec:results}
\begin{figure}
\centering
\includegraphics[width=0.9\textwidth, trim = 0cm 12.1cm 0cm 0cm, clip=true]{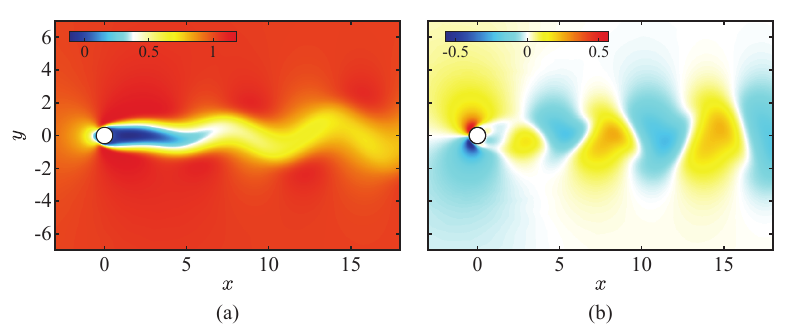}
\caption{A snapshot of the flow around a cylinder, oscillating on the limit cycle at $\Rey=50$. (a) $x$-component $u$ of the velocity field  (b) $y$-component $v$ of the velocity field.}
\label{fig:t12000}
\end{figure}

In this section, we exploit  
our approach to numerically obtain a reduced-order model and design a closed-loop controller for the flow around a cylinder. We consider the flow at $\Rey=50$, which is slightly above the critical value $\Rey_c$ at which the cylinder wake starts to oscillate. The proximity to $\Rey_c$ ensures that we are within the regime of validity of the weakly nonlinear analysis. Figure~\ref{fig:t12000} shows a snapshot of the velocity field obtained via DNS when the flow is fully saturated onto the limit cycle.

\subsection{Validation of the Stuart-Landau Model}
Here we present the construction of a Stuart-Landau model using the weakly nonlinear analysis from Section~\ref{sec:WNA}, with the coefficients $a_0$ and $a_1$ determined by~\eqref{eq:SLcoefficients}, and validate the model by comparing it to DNS results. Both the unforced and forced case with resonant forcing frequency $\omega_f=\omega_0$ are shown.

\begin{figure}
\centering
\includegraphics[width=0.9\textwidth,trim = 0cm 6.5cm 0cm 0cm,clip=true]{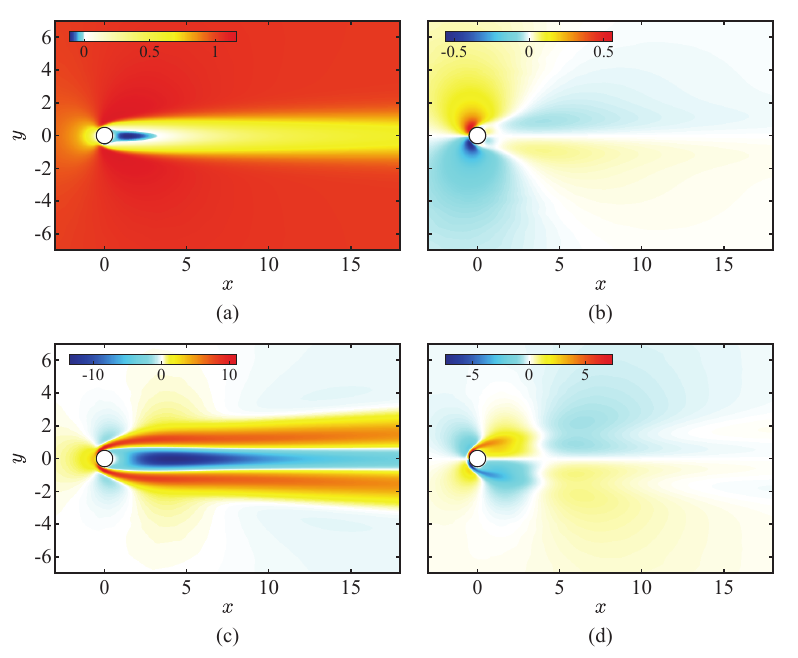}
\caption{Spatial structure of the base flow $\mathbf{u}_0$ at $Re_c=46.6$ and its first-order modification $\mathbf{u}_2^1$ due to an increase $\epsilon$ in Reynolds number, evaluated at $\Rey=50$. 
(a) The $x$-component $u_0$ of the base flow, (b) The $y$-component $v_0$ of the base flow, 
(c) The $x$-component $u_2^1$ of the base flow modification, (d) The $y$-component $v_2^1$ of the base flow modification.}
\label{fig:baseflow}
\end{figure}

\begin{figure}
\centering
\includegraphics[width=0.9\textwidth,trim = 0cm 6.5cm 0cm 0cm,clip=true]{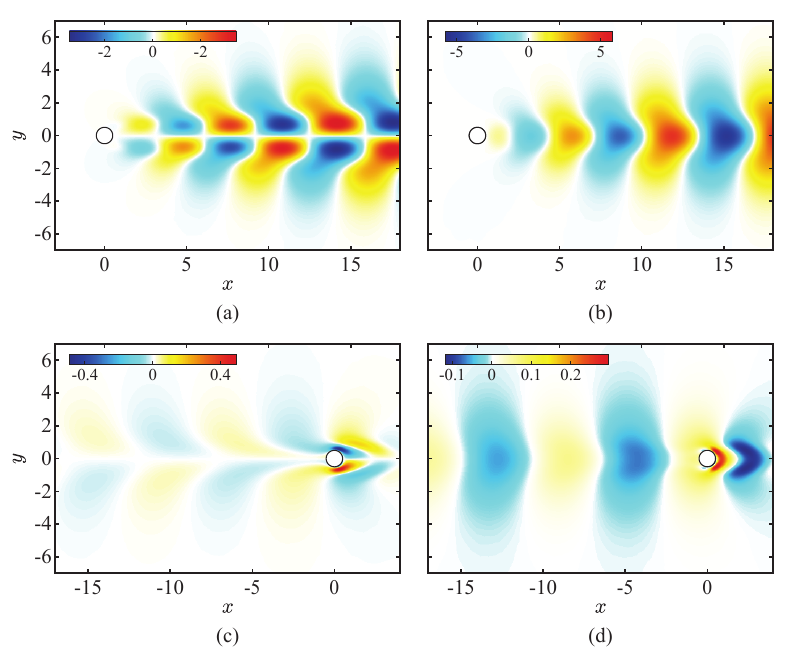}
\caption{Spatial structure of the global mode (first harmonic) $\mathbf{u}_1^A$ and the adjoint global mode $\mathbf{u}_1^{A \ast}$.
(a) Real part $\Re (u_1^{A})$ of the $x$-component of the global mode, (b) Real part $\Re (v_1^{A})$ of the $y$-component of the global mode, 
(c) Real part $\Re (u_1^{A \ast})$ of the $x$-component  of the adjoint global mode, (d) Real part $\Re (v_1^{A \ast})$ of the $y$-component of the adjoint global mode.}
\label{fig:globalmode}
\end{figure}

The nonlinear stationary Navier-Stokes equation~\eqref{eq:base} is numerically solved using an iterative Newton method to obtain the base flow $\mathbf{U}_0$ at $\Rey_c$, while the eigenvalue problem~\eqref{eq:eigenvalue} is solved via the shift-and-invert Arnoldi method as implemented in the ARPACK library (see~\cite{lehoucq1998arpack}). We obtain the eigenvalue $\lambda=0.0001+0.73741\imath$ at the critical Reynolds number $\Rey_c=46.6$, which has a negligible real part and is thus marginally stable. The velocity components of the base flow $\mathbf{u}_0$ at $\Rey_c=46.6$ and of the base flow modification $\mathbf{u}_2^1$ for $\Rey > \Rey_c$ are shown in Figure~\ref{fig:baseflow}. The region of negative $x$-component $u_0$ in Figure~\ref{fig:baseflow}(a) represents the steady recirculation region in the wake of the cylinder. The spatial extent of the recirculation region increases with the Reynolds number, as indicated by the negative value of $u_2^1$ in the wake in Figure~\ref{fig:baseflow}(c). In Figure~\ref{fig:globalmode}, the velocity components of the global mode $\mathbf{u}_1^A$ and the adjoint global mode $\mathbf{u}_1^{A\star}$ (note the displaced $x$-coordinate axis) are plotted. 
Plotting the real part of the transverse velocity component $v_1^A$ of the global mode, as in Figure~\ref{fig:globalmode}(b), reveals the oscillatory nature of the mode, which relates physically to the von Karman vortex street shown in Figure~\ref{fig:t12000}.

The coefficients of the Stuart-Landau model representing this flow are computed from~\eqref{eq:SLcoefficients} and given in Table~\ref{tab:SLcoeff}. Here, we normalized the global mode so that $v_1^A(1, 0) = 0.4612$. This results in $\Re(a_1) \approx \Re(a_0)$ and the magnitude of $A$ at the limit cycle being approximately $1$, as explained in~\cite{sipp2007global}. Both the mode shapes and model coefficients $a_0$ and $a_1$ are consistent with those in~\cite{sipp2007global}. The coefficient $a_2$ is calculated for the optimal forcing structure~\eqref{eq:optimal_forcing} and the localized forcing structure discussed in Section~\ref{sec:localforcing}.

\begin{table}
  \begin{center}
\def~{\hphantom{0}}
  \begin{tabular}{lcccccccc}
Approach & $a_0$   &  $a_1$  &  $a_2$-optimal forcing &  $a_2$-localized forcing \\ [5pt]
 WNA & $9.1219+3.2302\imath$ & $9.1053-31.1445\imath$ & $0.9939$ & $0.0942+0.002\imath$ \\
 LS & $8.672 + 5.2588 \imath$ & $5.901 - 14.5511 \imath$ & 
 $0.8106 - 0.1481 \imath$ & $0.1234 - 0.0066 \imath$
  \end{tabular}
   \caption{Values of the coefficients of the Stuart-Landau model obtained via weakly nonlinear analysis (WNA) and least-squares regression (LS) to DNS data. The coefficient $a_2$ is
calculated for the optimal forcing structure~\eqref{eq:optimal_forcing} and the localized forcing structure~\eqref{eq:localforcing}.}
   \label{tab:SLcoeff}
  \end{center}
\end{table}
For model validation purposes, direct numerical simulation is performed to obtain the exact solution $\mathbf{U}=\mathbf{U}_b+\mathbf{U}'$ of the incompressible Navier-Stokes equations~\eqref{eq:NS_matrix1} at $\Rey=50$, for which $\epsilon=0.0015$.  
First, we calculate the base flow $\mathbf{U}_b$  using the iterative Newton method. The unsteady perturbation field $\mathbf{U}'$ is then marched in time as explained in Section~\ref{sec:DNS}. We validate our model by comparing the evolution of the global mode amplitude $A$ predicted by the model with the $\tilde{A}$ evaluated from the DNS results via expression~\eqref{eq:Atilde}.
The difference $e_A$ between $A$ and $\tilde{A}$ is
\begin{align} \label{eq:DNS_projected_resonant} 
   e_A \equiv \tilde{A}- A:= \frac{1}{\sqrt{\epsilon}} \langle \mathbf{u}_1^{A \star}, \mathbf{e}_{\mathbf{u}}\rangle e^{-\imath \omega_0 t},
\end{align}
where the component of the approximation error associated with the velocity field is
\begin{align}\label{eq:approximation_error}
\mathbf{e}_{\mathbf{u}}=\mathbf{u} - ( \mathbf{u}_0+\sqrt{\epsilon} (Ae^{\imath \omega_0t}\mathbf{u}_1^A+c.c.) +\epsilon \mathbf{u}_2^1).
\end{align}

\begin{figure}
\centering
\includegraphics[width=0.9\textwidth,trim = 0cm 12.4cm 0cm 0cm,clip=true]{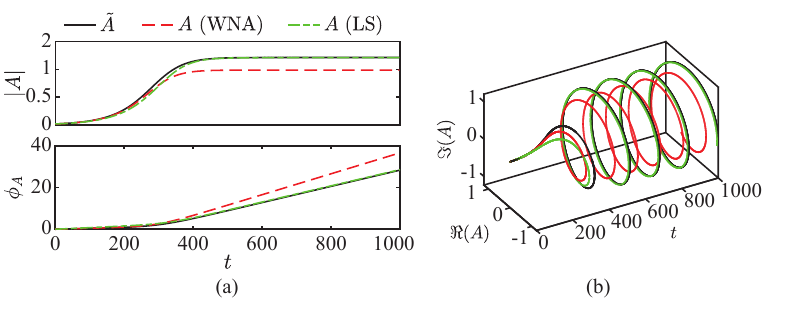}
\caption{Prediction of the time evolution of the global mode amplitude $A$ without applied forcing ($\mathbf{f}(t)=0$), at $\Rey=50$. The values of $A$ are estimated as $\tilde{A}$ from DNS assuming complete knowledge of the velocity field within the entire computational domain, and obtained from the Stuart-Landau model with coefficients derived from both our weakly nonlinear analysis (WNA) and a least-squares regression (LS) to the DNS estimate. $A$ is plotted in the polar form $A=|A|e^{i \phi_A}$: (a) Magnitude $|A|$ (top) and  phase $\phi_A$ (bottom), and (b) phase portrait.}
\label{fig:A_uncontrolled} 
\end{figure}

First, we examine the unforced case where $\mathbf{f}(t)=0$. The evolution of the global mode in this case is shown in Figure~\ref{fig:A_uncontrolled}. The perturbation velocity is initialized in the subspace of the global mode. Therefore, $\mathbf{e}_{\mathbf{u}}(t=0)=0$ for the approximation error in~\eqref{eq:approximation_error} and $A(t=0)=\tilde{A}(t=0)=0.019$. It can be seen that the accuracy is high around the base flow and it deteriorates as the limit cycle is approached. The unforced Stuart-Landau model underpredicts the magnitude of $A$ at the limit cycle by $\frac{|e_A|}{|\tilde{A}(t=500)|}=0.18 (18\%)$, although 
the general trend of the time evolution of $A$ is accurately predicted by the model.

To validate the forced Stuart-Landau model for the resonant forcing frequency $\omega_f=\omega_0$, we consider a slowly time-varying complex forcing amplitude $E(\tau)=\Re(E(\tau))+\imath \Im(E(\tau))$.  In particular, we consider both $\Re(E(\tau))$ and $\Im(E(\tau))$ in the form of a Schroeder Phased Harmonic Sequence (SPHS)~(\cite{schroeder1970synthesis}), which is commonly used for system identification purposes and is given as
\begin{equation}\label{eq:sphs}
    \Re / \Im (E(\tau)):=P\sum_{k=1}^{N}\sqrt{\frac{2}{N}} \cos\Big( \frac{2\pi k}{T}(\tau)+\theta_k\Big).
\end{equation}
Here $N$ is the number of harmonics contained in the 
signal, $P$ is the amplitude of all harmonics, $T$ is the period of the fundamental 
wave, and $\theta_k$ is the initial phase of the $k$th harmonic which we set to be different for $\Re(E(\tau))$ and $\Im (E(\tau))$. 

\begin{figure}
\centering
\includegraphics[width=0.9\textwidth,trim = 0cm 10cm 0cm 0cm,clip=true]{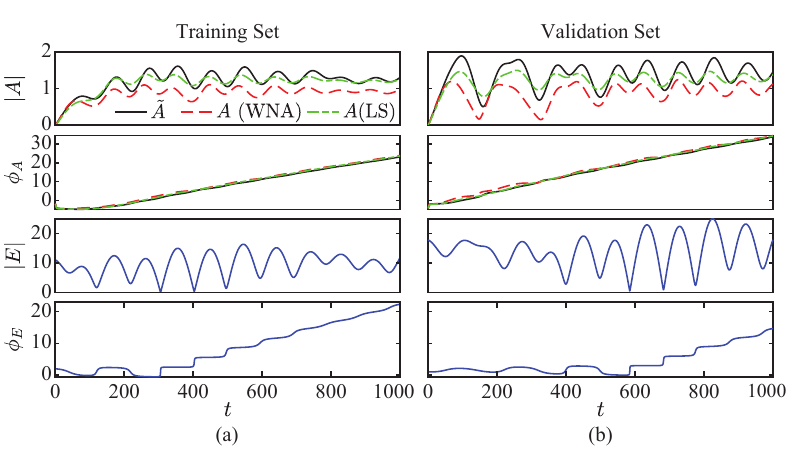}
\caption{Prediction of the time evolution of the global mode amplitude $A$ under resonant forcing $\omega_f=\omega_0$ with optimal spatial structure~\eqref{eq:optimal_forcing} at $\Rey=50$. The initial condition used for the DNS is the unforced solution at $t=40$ and the forcing amplitude $E(\tau)$ is of the form of Schroeder Phased Harmonic Sequence~\eqref{eq:sphs} with $N=20$ and $T=5$. Both $A$ and $E$ are plotted in polar form, $A=|A|e^{\imath \phi_A}$ and $E=|E|e^{\imath \phi_E}$. $A$ is estimated as $\tilde{A}$ from DNS assuming complete knowledge of the velocity field within the computational domain, and obtained from the Stuart-Landau model with coefficients derived from both our weakly nonlinear analysis and a least-squares regression to the DNS estimate. (a) Case with forcing amplitude $P=7.35$, which was used as a training set for the least-squares regression. (b) Case with forcing amplitude $P=10.5$, used to validate the least-squares regression.}
\label{fig:A_forced}
\end{figure}

In Figure~\ref{fig:A_forced}, we compare the evolution of the global mode predicted by the forced Stuart-Landau model~\eqref{eq:SL_resonance} to the DNS results for two forcing amplitudes $E$ of the form~\eqref{eq:sphs}, which differ in their amplitudes $P$ and initial phases $\theta_k$, while their other parameters are the same. We see that our Stuart-Landau model captures the qualitative behaviour of $A$ well for both cases. As in the unforced case, the accuracy is high around the base flow, while at the limit cycle the model underpredicts the magnitude of $A$. The accuracy of the model is overall lower for the stronger forcing in Figure~\ref{fig:A_forced}(b), which has a higher value of $P$, compared to the weaker forcing in Figure~\ref{fig:A_forced}(a). This is possibly related to the more prominent transient modes activated by the forcing.

\subsection{ Stuart-Landau Model with Data-Driven Coefficients }

To improve the accuracy of the Stuart-Landau model in predicting the magnitude of $A$ at the limit cycle when its coefficients are obtained from the weakly nonlinear analysis, 
we additionally explore an alternative, data-driven approach where we fit the coefficients of the Stuart-Landau equation to DNS data. 
The coefficients $a_0$ and $a_1$, given in Table~\ref{tab:SLcoeff}, are calculated via the least-squares regression of the time series of $\tilde{A}$ obtained from the DNS, which is shown in Figure~\ref{fig:A_uncontrolled}. We use the time window $t=[0,600]$ for the fitting. We can see from Figure~\ref{fig:A_uncontrolled}, that the Stuart-Landau model equipped with the regressed coefficients predicts the behavior around the limit cycle better than when using the coefficients derived from the weakly nonlinear analysis. However, accuracy is sacrificed in the prediction of the growth from the base flow to the limit cycle.

For the forced case, we retain the coefficients $a_0$ and $a_1$ obtained from fitting the unforced DNS data and we fit coefficient $a_2$ (given in Table~\ref{tab:SLcoeff}) to the forced DNS data. In the case of an optimal forcing structure, we use the results from Figure~\ref{fig:A_forced}(a) as the training set. The forcing amplitude $P$ is chosen such that the maximum magnitude of $E$ is close to the one used later during closed-loop control (see Figure~\ref{fig:control_opt}(b)). Unlike the model given by the weakly nonlinear analysis, this model is not inherently parametrized as it is fitted at a fixed parameter value. The model is validated on a data set with a higher forcing amplitude, shown in Figure~\ref{fig:A_forced}(b). As in the unforced case, we see an improvement in accuracy around the limit cycle and a slight deterioration in the region of growth from the base flow to the limit cycle, compared to the behavior predicted by the Stuart-Landau model with coefficients obtained from the weakly nonlinear analysis.

\subsection{Closed-Loop Control}
Here we apply our control design from Section \ref{sec:control} to suppress the cylinder wake oscillation at $\Rey=50$. First, we apply a forcing with an optimal spatial distribution and estimate the reduced state based on knowledge of the velocity field within the spatial bounds of the computational domain, which we refer to as full domain measurements. Then we introduce a spatially compact forcing structure, which is more physically realizable than the distributed optimal volume forcing. We also determine suitable locations to measure the velocity, 
which enable the effective estimation of the reduced state. These choices represent a more physically realizable system, since sufficiently fast full-domain measurements may be difficult or impossible to obtain. We perform the closed-loop control using the selected compact forcing and measurement locations, which enable the real-time implementation of the design.

\subsubsection{Control with optimal forcing structure and full measurements}\label{sec:controloptimal}
As a starting point, we consider the ideal case for the control design where resonant forcing with optimal spatial distribution $\mathbf{f}_E$~\eqref{eq:optimal_forcing} and full domain measurements are assumed. 

In our implementation, the controller is switched on once the flow is settled on the limit cycle. The time-discretized sequence of control inputs $\mathbf{q_j}=[\Re(E_j), \Im(E_j)]^T$ is determined by the model predictive control scheme in Section~\ref{sec:mpc}, which minimizes the cost function~\eqref{eq:costfunction}, and implemented by the continuous-time system using a zero-order hold. To solve the minimization problem, the prediction of the future states $\mathbf{X}_{j+k}=[\Re(A_{j+k}), \Im(A_{j+k})]^T$ is done by numerical integration of the Stuart-Landau model~\eqref{eq:SL_resonance_t}. Here we use the Stuart-Landau model with coefficients obtained both from the weakly nonlinear analysis and the least-squares regression to the DNS data.

\begin{figure}
\centering
\includegraphics[width=0.9\textwidth+30.295pt,trim = 0cm 9cm -33.6612pt 0cm,clip=true]{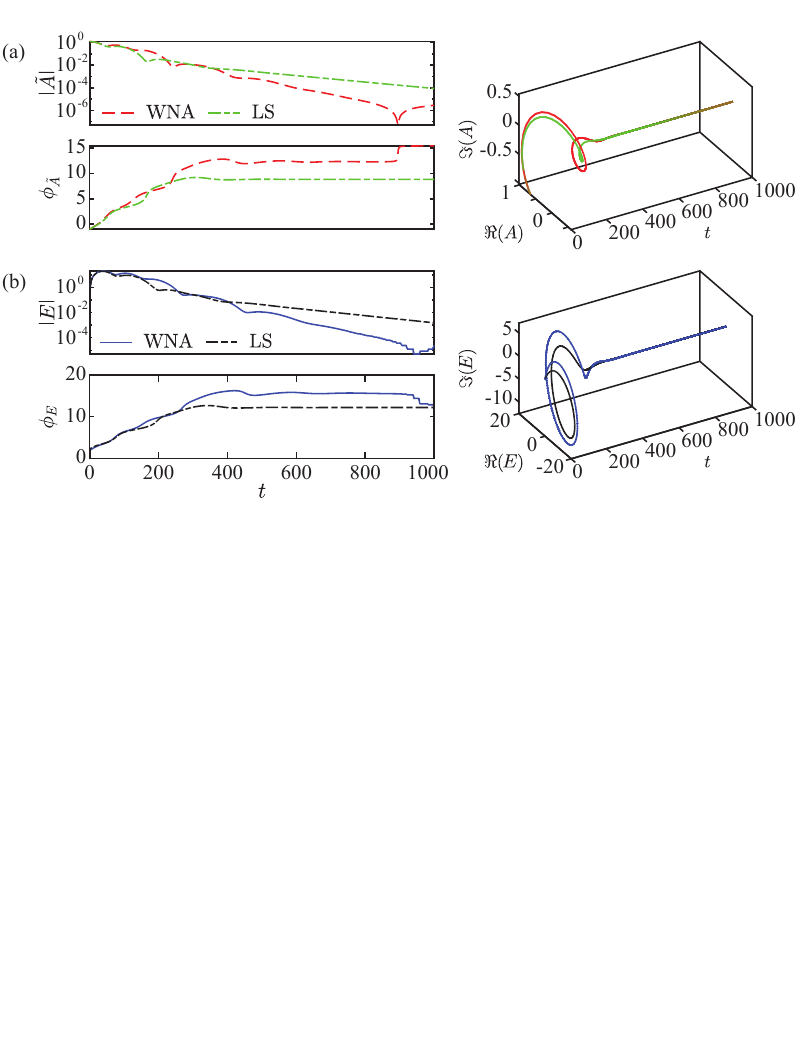}
\caption{Suppression of the global mode amplitude $A$ with resonant forcing of amplitude $E$ and optimal forcing structure $\mathbf{f}_E$~\eqref{eq:optimal_forcing} using MPC. Results using the Stuart-Landau model with coefficients obtained both from the weakly nonlinear analysis and from the least-squares regression are shown. The reduced state $\tilde{A}$ is estimated from the full measurement of the velocity field obtained with DNS. (a) Time history of $\tilde{A}$. Magnitude $|\tilde{A}|$ and phase $\phi_{\tilde{A}}$ (left) and phase portrait (right) (b) Time history of $E$. Magnitude $|E|$ and phase $\phi_{E}$ (left) and phase portrait (right)}
\label{fig:control_opt}
\end{figure}

\begin{figure}
\centering
\includegraphics[width=0.9\textwidth, trim = 0cm 12.6cm 0cm 0cm,clip=true]{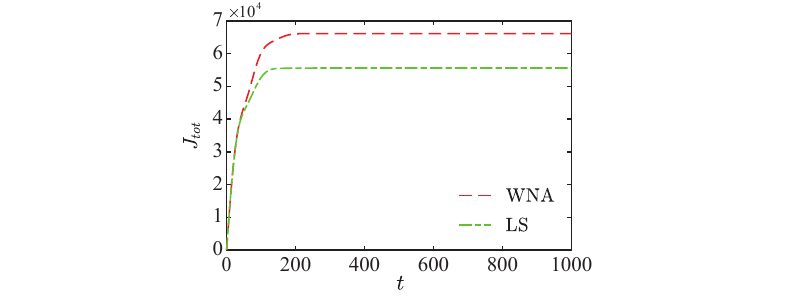}
\caption{ Cumulative cost $J_{\rm tot}(n)$ over time $t=n\Delta t$ ($\Delta t=1$) of the controller based on the Stuart-Landau model with coefficients derived from the weakly nonlinear analysis versus that with coefficients obtained via the least-squares regression. Optimal forcing structure~\eqref{eq:optimal_forcing} and full measurements are used for the controller design.}
\label{fig:cumulative_cost_opt}
\end{figure}

The reduced state $\tilde{\mathbf{X}}_j=[\Re(\tilde{A}_{j}), \Im(\tilde{A}_{j})]^T$ is estimated according to~\eqref{eq:Atilde} from the velocity field obtained via DNS.
We use the sampling time step $\Delta t=1$ and
the weighting matrix $\mathbf{R}_{\Delta u}=8 \mathbf{I}$ to ensure small temporal gradients of $E$ and thus comply with the assumption of its slow variation. The weighting matrices $\mathbf{Q}$ and $\mathbf{R}$ are set to $1000\mathbf{I}$ and $0.9 \mathbf{I}$, and the horizon of $m=5$ time steps is used. We ran our simulations for $1000$ time steps, which is sufficient to bring $|\tilde{A}|$ and $|E|$ to very small values for both models. From Figure~\ref{fig:control_opt}, we see that using the model obtained by the weakly nonlinear analysis, we bring $|\tilde{A}|$ and $|E|$ closer to zero than using the data-driven counterpart. To further compare the performance of the controllers based on each of the two models, we calculate the cumulative cost
\begin{equation}
  J_{\rm tot}(n):= \sum_{j=1}^{n}\| \tilde{\mathbf{X}}_{j+1}-\mathbf{X}^{\ast} \|^2_{\mathbf{Q}}+ \sum_{j=1}^{n}\big( \| \mathbf{q}_{j}\|^2_{\mathbf{R_u}}+ \| \Delta \mathbf{q}_{j}\|^2_{\mathbf{R_{\Delta u}}}\big).
\end{equation}
where $n$ denotes the number of time steps of the simulation.
The cumulative cost is smaller for the  Stuart-Landau model obtained via the least-squares regression as shown in  Figure~\ref{fig:cumulative_cost_opt}. This is justified by the fact that the least-squares model is more accurate around the limit cycle, which is the starting point of our simulation, and where the control cost is significantly higher. Thus, due to the restricted length of the simulation, the least-squares model yields a smaller cumulative cost. However, the accuracy of the Stuart-Landau model obtained from the weakly nonlinear analysis is higher away from the limit cycle. This explains why smaller values of $|\tilde{A}|$ and $|E|$ are reached with this model as it converges to the equilibrium.

\begin{figure}
\centering
\includegraphics[width=0.9\textwidth, trim = 0cm 12.6cm 0cm 0cm,clip=true]{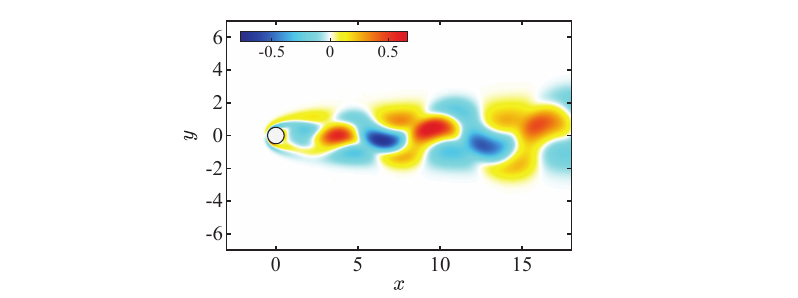}
\caption{Snapshot of perturbation vorticity $\mathbf{\omega}'$ at the limit cycle. }\label{fig:vorticity_uncontrolled}
\end{figure}

\begin{figure}
\centering
\includegraphics[width=0.9\textwidth, trim = 0cm 12.1cm 0cm 0cm,clip=true]{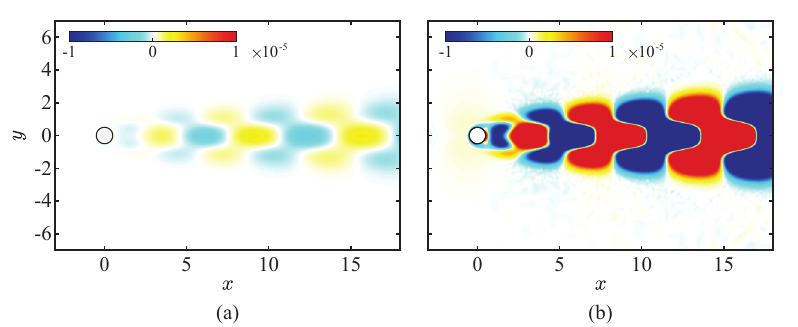}
\caption{Perturbation vorticity $\mathbf{\omega}'$ at the final time $t=1000$ of the flow controlled with the optimal forcing structure~\ref{eq:optimal_forcing} and using full measurements. (a) Stuart-Landau model with coefficients derived from the weakly nonlinear analysis (b) Stuart-Landau model with coefficients obtained via the least-squares regression. }\label{fig:vorticity_controlled_opt}
\end{figure}

In both cases, the control objective, which is to suppress the unsteadiness of the flow, is achieved. The suppression of unsteadiness is evaluated by plotting the perturbation vorticity $\mathbf{\omega}'=\nabla \times \mathbf{u}'$ of the uncontrolled flow at the limit cycle (Figure~\ref{fig:vorticity_uncontrolled}) and comparing it to the controlled case at the final time $t=1000$. From Figure~\ref{fig:vorticity_controlled_opt}, we can see that for both models, the unsteadiness of the flow is almost fully suppressed. In analogy to the decay in $|A|$ and $|E|$, the residual vorticity is higher for the model with data-driven coefficients by two orders of magnitude.


\subsubsection{Localized Forcing Structure}\label{sec:localforcing}

\begin{figure}
\centering
\includegraphics[width=0.9\textwidth, trim = -0.1cm 12.6cm 0.1cm 0cm,clip=true]{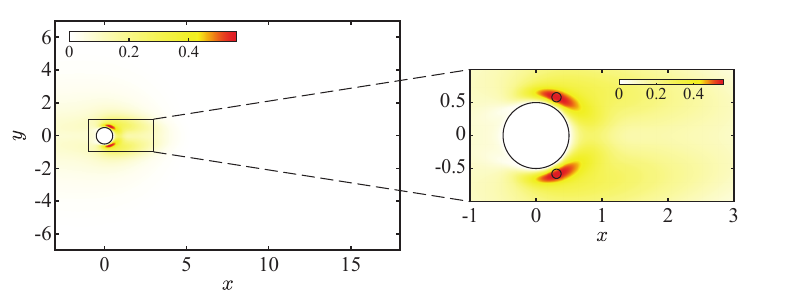}
\caption{Spatial distribution of the norm of the adjoint mode velocity $||\mathbf{u}_1^{A \star}(\mathbf{x})||$. Circles indicate our choice of localized forcing domain $\Omega_s$, with chosen radius $0.07$ and centered on locations $\mathbf{x}_i=(0.3, \pm 0.58)$ where this norm is the highest, thus representing favorable locations for forcing.}
\label{fig:a2norm}
\end{figure}

The previously presented results were obtained with the optimal forcing structure, which is widely spatially distributed according to the adjoint eigenvector shown in Figure~\ref{fig:globalmode}(c) and Figure~\ref{fig:globalmode}(d). Applying a widely spatially distributed forcing is however unfeasible for real applications. Here, we consider a more physically realisable forcing structure concentrated on a small domain $\Omega_s$. From~\eqref{eq:SLcoefficients} we know that the structure $\mathbf{f}_E$ which maximizes $|a_2|$ is in the direction of $\mathbf{u}_1^{A \star}$ on $\Omega_s$. For a sufficiently small domain $\Omega_s$, we can assume $\mathbf{u}_1^{A \star}(\mathbf{x})$ to be approximately constant on $\Omega_s$. It follows that $|a_2|=||\mathbf{u}_1^{A \star}(\mathbf{x}_i)|| \sqrt{\rm vol(\Omega_s)}$ for uniform, normalized $\mathbf{f}_E$, where $ ||\mathbf{u}_1^{A \star}(\mathbf{x}_i)||=\sqrt{|u_1^{A \star}(\mathbf{x}_i)|^2+|v_1^{A \star}(\mathbf{x}_i)|^2}$ is the norm of $\mathbf{u}_1^{A \star}$ evaluated at the spatial location $\mathbf{x}_i$. This indicates that to maximize $|a_2|$, the volume force should be applied at location $\mathbf{x}_i$ where the norm of the adjoint mode is the highest. From Figure~\ref{fig:a2norm}, we can see that this location is approximately at $\mathbf{x}_i=(x_i, y_i)=(0.3, \pm 0.58)$. Thus, we choose a forcing structure as 
\begin{equation}\label{eq:localforcing}
  \mathbf{f}_E:=\frac{1}{\sqrt{{\rm vol}(\Omega_{s_{+}}\cup\Omega_{s_{-}})}}\Big(\frac{\mathbf{u}_1^{A \star}(0.3, 0.58)}{||\mathbf{u}_1^{A \star}(0.3,0.58)||}\chi_{\Omega_{s_+}}+\frac{\mathbf{u}_1^{A \star}(0.3, -0.58)}{||\mathbf{u}_1^{A \star}(0.3,-0.58)||}\chi_{\Omega_{s_-}} \Big),  
\end{equation}
where $\chi_S$ denotes the indicator function of a set $S\subset\mathbb R^2$, which takes the value $1$ on $S$ and $0$ outside. We choose $\Omega_{s_{\pm}}$ to be circular domains with a radius of $0.07$ and center at $(0.3, \pm 0.58)$, indicated with the circles in Figure~\ref{fig:a2norm}. 

\subsubsection{Point velocity measurements}

In the results presented earlier, we used the state estimated from the velocity field obtained with DNS to design our feedback. However, measuring the velocity field over a large domain using, for example, methods such as particle image velocimetry (PIV), is generally too slow for real-time control.
In real applications the flow is instead usually measured using a limited number of spatially discrete sensors, which typically capture only partial information about the velocity field. Therefore, we assume the existence of sensors that provide point measurements of the velocity.
The goal is to calculate $\tilde{A}_p$ via mapping~\eqref{eq:measurements}, which closely approximates $\tilde{A}$ obtained from the full measurements with mapping~\eqref{eq:Atilde}, using as few point velocity measurements $\mathbf{x}_1,\dots,\mathbf x_n$ as possible. In analogy to~\eqref{eq:error1}, we define the approximation error
\begin{equation}\label{eq:error} \tilde{\mathbf{e}}\equiv(\tilde{\mathbf{e}}_{\mathbf{u}}, \tilde{e}_p):=\mathbf U-\mathscr{G}(\tilde{\mathbf{X}})
\end{equation}
where $\mathscr{G}$ is given by~\eqref{eq:Gmap}.
If $\tilde{\mathbf{e}}_{\mathbf{u}}=0$ mappings~\eqref{eq:Atilde} and~\eqref{eq:measurements} should give the same estimate of $A$, i.e. $\tilde{A}=\tilde{A}_p$ when the matrix in~\eqref{eq:measurements} has full-column rank. Therefore, appropriate locations $\mathbf{x}_1,\ldots,\mathbf x_n$ for which $\tilde{A}_p$ in~\eqref{eq:measurements} closely approximates $\tilde{A}$ are  locations where the error $\tilde{\mathbf{e}}_{\mathbf{u}}$ is small and the magnitude of the global mode is sufficiently large. 

\begin{figure}
\centering
\includegraphics[width=0.9\textwidth, trim = -0.1cm 12.6cm 0.1cm 0cm,clip=true]{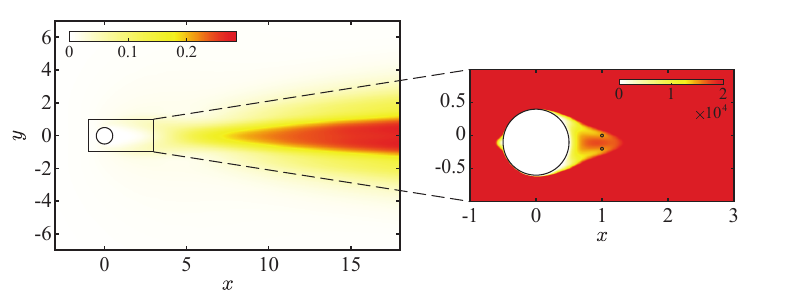}
\caption{Spatial distribution of the norm of the root mean-square error $ ||\tilde{\mathbf{e}}_{\mathbf{u}_{\rm rms}}(\mathbf{x}_i)||=\sqrt{|\tilde{e}_{u_{{\rm rms}}}(\mathbf{x}_i)|^2+|\tilde{e}_{v_{\rm rms}}(\mathbf{x}_i)|^2}$ . Circles denote suitable locations $\mathbf{x}_1=(1,0.1)$ and  $\mathbf{x}_2=(1,-0.1)$ chosen for taking point measurements of the velocity, where $||\tilde{\mathbf{e}}_{\mathbf{u}_{\rm rms}}(\mathbf{x}_i)||$ is small and $\mathbf{u}_1^A$ is non-negligible.} \label{fig:error}
\end{figure}

\begin{figure}
\centering
\includegraphics[width=0.9\textwidth,trim = 0cm 12.4cm 0cm 0cm,clip=true]{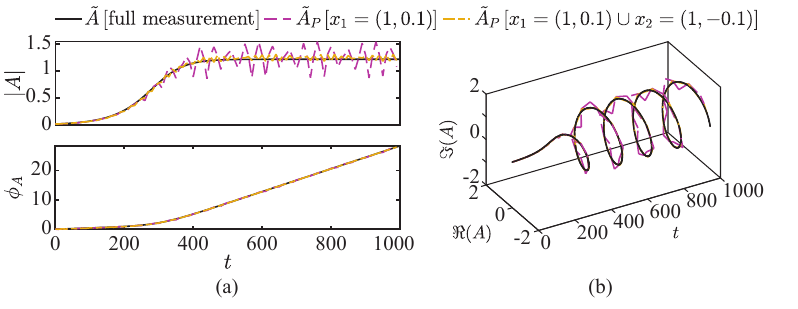}
\caption{Time evolution of the global mode amplitude $\tilde{A}$ (without applied forcing) estimated from full measurement of the velocity field vs. $\tilde{A}_p$ estimated from a single point measurement at $\mathbf{x}_1=(1,0.1)$ and two-point measurement at $\mathbf{x}_1=(1,0.1)$ and $\mathbf{x}_2=(1,-0.1)$. $\tilde{A}$ and $\tilde{A}_p$ are plotted in the polar form: (a) Magnitude (top) and phase (bottom), and (b) phase portrait.}
\label{fig:A_meas}
\end{figure}

To find these locations, we analyse the root mean square error $\tilde{\mathbf{e}}_{\mathbf{u}_{\rm rms}}=(\tilde{e}_{u_{\rm rms}}, \tilde{e}_{v_{\rm rms}})$ 
over one period of the limit cycle of the unforced flow, where the approximation error is the highest. Figure~\ref{fig:error} shows the spatial distribution of the norm of this root mean square error, i.e. $ ||\tilde{\mathbf{e}}_{\mathbf{u}_{\rm rms}}(\mathbf{x})||=\sqrt{|\tilde{e}_{u_{\rm rms}}(\mathbf{x})|^2+|\tilde{e}_{v_{\rm rms}}(\mathbf{x})|^2}$ across the spatial location $\mathbf x$.
We select the location $\mathbf{x}=(1,0.1)$ at which the error is small and the global mode has a considerable magnitude (see Figure~\ref{fig:globalmode})
and compare $\tilde{A}_p$ against $\tilde{A}$ in Figure~\ref{fig:A_meas}. We can see that the oscillatory discrepancy when estimating $\tilde{A}_p$ grows as the flow is approaching the limit cycle due to the growth of the approximation error. This is reduced by adding an additional measuring point symmetrically with respect to axis $y$. 

\subsubsection{Control with localized forcing structure and point velocity measurement}
Finally, we present the results of applying closed-loop control with forcing of localized structure, as presented in Section~\ref{sec:localforcing}, and assuming we can measure the velocity at two points. We choose the points $\mathbf{x}_1=(1,0.1)$ and $\mathbf{x}_2=(1,-0.1)$ which were shown to be suitable locations for accurate estimation of the reduced state. 

\begin{figure}
\centering
\includegraphics[width=0.9\textwidth+30.295pt,trim = 0cm 9cm -33.6612pt 0cm,clip=true]{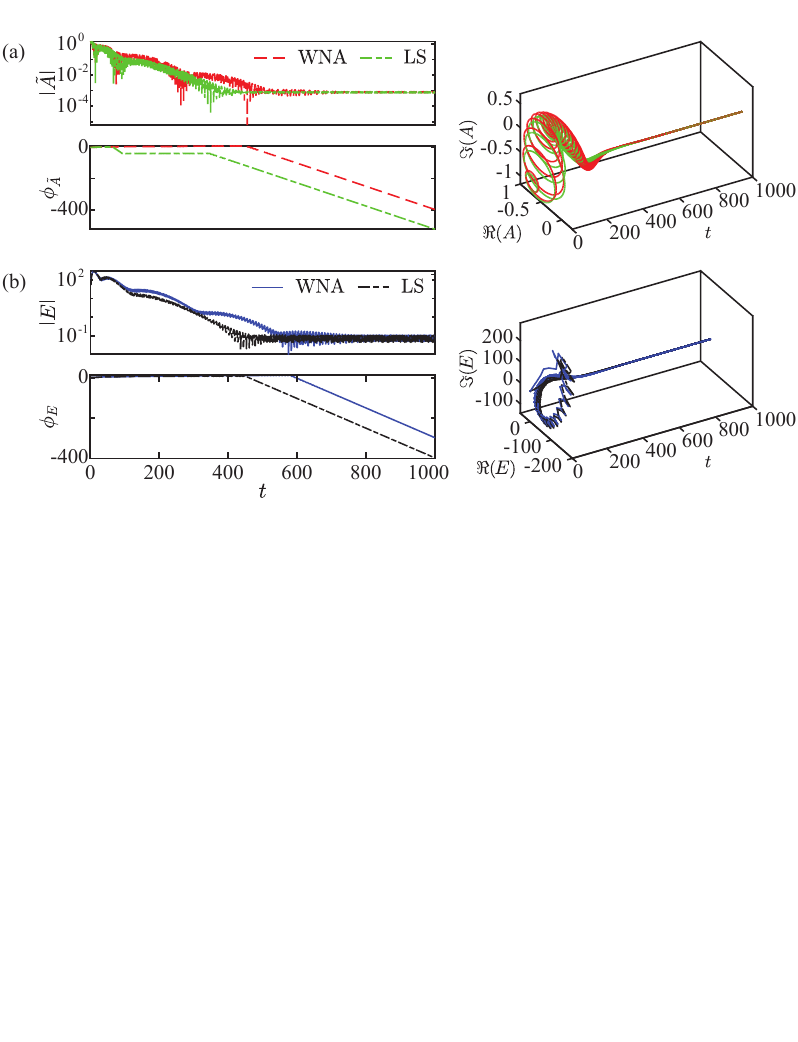}
\caption{Suppression of the global mode amplitude $A$ with resonant forcing of amplitude $E$ and localized forcing structure $\mathbf{f}_E$~\eqref{eq:localforcing} using MPC. Results using the Stuart-Landau model with coefficients obtained both from the weakly nonlinear analysis and from the least-squares regression are shown. The reduced state $\tilde{A}$ is estimated from the two-point velocity measurements at $\mathbf{x}_1=(1,0.1)$ and $\mathbf{x}_2=(1,-0.1)$. (a) Time history of $\tilde{A}$. Magnitude $|\tilde{A}|$ and phase $\phi_{\tilde{A}}$ (left) and phase portrait (right) (b) Time history of $E$. Magnitude $|E|$ and phase $\phi_{E}$ (left) and phase portrait (right).}
\label{fig:control_nonopt}
\end{figure}

\begin{figure}
\centering
\includegraphics[width=0.9\textwidth, trim = 0cm 12.6cm 0cm 0cm,clip=true]{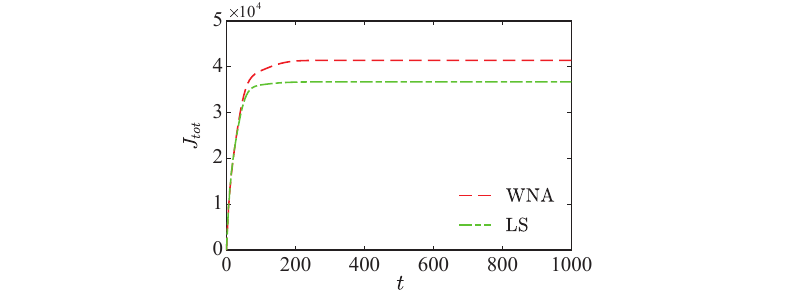}
\caption{Cumulative cost $J_{tot}$ over time of the controller based on the Stuart-Landau model with coefficients derived from the weakly nonlinear analysis versus that with coefficients obtained via the least-squares regression. Localized forcing structure~\eqref{eq:localforcing} and two-point velocity measurements at $\mathbf{x}_1=(1,0.1)$ and $\mathbf{x}_2=(1,-0.1)$ are used for the controller design.}\label{fig:cumulative_cost_meas}
\end{figure}

The weighting matrices for MPC are set as: $\mathbf{Q}= 1000\mathbf{I}$, $\mathbf{R}=0.1 \mathbf{I}$ and $\mathbf{R}_{\Delta u}=0.001 \mathbf{I}$ while the rest of the settings are the same as in Section~\ref{sec:controloptimal}  We run our simulation until $t=1000$. From Figure~\ref{fig:control_nonopt}, we see wiggling of the state $\tilde{A}$ which is due to the oscillatory discrepancy present when estimating it from point measurements.  Since an additional error is introduced, we get similar final decay in $|\tilde{A}|$ and $|E|$ for both models. This decay is, as expected, less pronounced than observed in the optimal case, however the Stuart-Landau model with coefficients obtained by least-squares regression again shows smaller cumulative cost (see Figure~\ref{fig:cumulative_cost_meas}). 

\begin{figure}
\centering
\includegraphics[width=0.9\textwidth, trim = 0cm 12.1cm 0cm 0cm,clip=true]{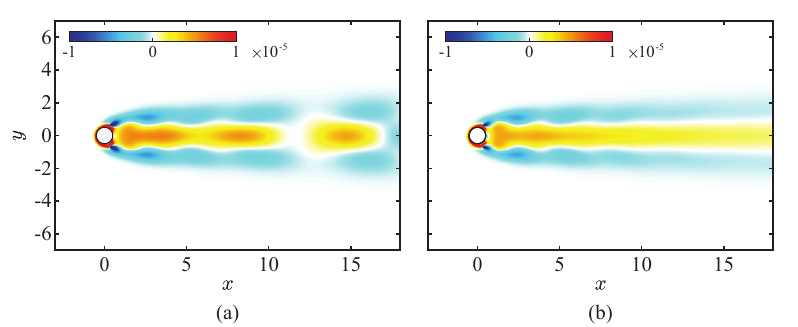}
\caption{Perturbation vorticity $\mathbf{\omega}'$ at the final time $t=1000$ of the flow controlled with the localized forcing structure~\eqref{eq:localforcing} and using two-point velocity measurements at $\mathbf{x}_1=(1,0.1)$ and $\mathbf{x}_2=(1,-0.1)$. (a) Stuart-Landau model with coefficients derived from the weakly nonlinear analysis (b) Stuart-Landau model with coefficients obtained via the least-squares regression. }\label{fig:vorticity_meas}
\end{figure}

The perturbation vorticity at the final time step is shown in Figure~\ref{fig:vorticity_meas}, from which we see that the unsteadiness is again almost fully suppressed. However, compared to the optimal case, the perturbation vorticity around the cylinder is higher.

\section{Conclusion}
In this work, we presented a methodology for model-based closed-loop control of vortex shedding in the
cylinder wake. This methodology is applicable to any fluid flow that is described by the incompressible Navier-Stokes equations and undergoes its first Hopf bifurcation.
We derived a parametric reduced-order model of the incompressible Navier-Stokes equations in the form of a forced Stuart-Landau equation, which governs the evolution of a global mode amplitude $A(\tau)$ on a slow time scale $\tau$. The model, together with corresponding mappings from the low-dimensional state and input space to the full state and input space, are obtained via a global weakly nonlinear analysis around the critical Reynolds number at which a supercritical Hopf bifurcation occurs. To design a feedback law based on this model, we applied a harmonic volume forcing $\mathbf{f}=\sqrt{\epsilon}^i E(\tau)e^{\imath \omega_f t}\mathbf{f}_E+c.c.$ with the amplitude $E(\tau)$ varying on the slow time scale. We showed that different classes of forcing frequencies require different scaling $\sqrt{\epsilon}^i$ and lead to different forms of the Stuart-Landau equation, which is in accordance with the results in~\cite{sipp2012open} for constant $E$. The approximation of the full state, i.e., the velocity-pressure field $\mathbf{U}$, by the weakly nonlinear analysis, showed that the flow is at its steady state when both $A$ and $E$ are zero. Controlling the flow near the resonant frequency $\omega_0$ or its half $\frac{\omega_0}{2}$, allowed us to bring both $A$ and $E$ to zero. This enables us to fully suppress the unsteadiness, which is not possible for the rest of the forcing frequencies. Among the two forcing frequencies that achieve full suppression, we selected $\omega_f \approx \omega_0$, which facilitates determining the optimal forcing structure.

We proposed a three-step output feedback control design using velocity measurements. First, we estimated the reduced state from partial velocity measurements and fed this estimate to the controller. The controller then determined the control amplitude $E$ which brings $A$ to zero based on the surrogate model for $\omega_f \approx \omega_0$. This was done via a model predictive controller, which allowed us to include soft constraints on the rate of change of the forcing amplitude and its magnitude. Thus we were able to comply with our modelling requirement of small temporal gradients of $E$ and bring $E$ to zero. 
Finally, we mapped the control input of the reduced-order model back into the forcing $\mathbf{f}$ of the true system. We showed that the optimal forcing structure at the resonant frequency has the shape of the adjoint global mode. In addition, we presented a simple and efficient way to estimate the reduced state from a finite number of measurements. 
 

Applying our methodology to control the cylinder wake at $\Rey=50$, we saw that the model predicts the general trend of the evolution of $A$ accurately but underpredicts its magnitude at the limit cycle. Thus, we additionally introduced a data-driven approach where we fitted the
coefficients of the Stuart-Landau equation to DNS data. We saw that this helps increase the accuracy of the reduced-order model at the limit cycle but deteriorates the accuracy in predicting the growth from the base flow to the limit cycle. These modifications, however, reduced the overall cost to control the system. 
We also determined optimal locations to apply a spatially dense, localized volume force, as well as suitable locations to place measurement sensors. By choosing two locations in the near wake of the cylinder, which are symmetric around $y$, we accurately evaluated the amplitude of the global model. 
%
%
Finally, we successfully demonstrated full suppression of the oscillations in the cylinder wake using spatially dense, localized volume forcing and two-point velocity measurements.






\bibliographystyle{IEEEtranS}
\bibliography{jfm-instructions}

\end{document}